\documentclass[superscriptaddress,10pt,twocolumn,notitlepage]{revtex4-1}
 \usepackage{multirow}
 \usepackage{mathptmx}
\usepackage{graphicx}
\usepackage[dvipsnames]{xcolor}
\usepackage{amsmath}
\usepackage{amssymb}
\usepackage{amscd}
\usepackage{bm}
\usepackage{enumerate}
\usepackage{type1cm}
\usepackage{lettrine}
\usepackage{mathrsfs}
\usepackage[displaymath,mathlines]{lineno}
\usepackage[space]{grffile}
\usepackage{calrsfs}
\usepackage{epsfig}
\usepackage{subfigure}
\usepackage{psfrag}
\usepackage[T1]{fontenc}

\usepackage{color}
\usepackage{placeins}
\usepackage{wrapfig}
 
\def\beq{\begin{equation}}
\def\eeq{\end{equation}}
\def\bea{\begin{eqnarray}}
\def\eea{\end{eqnarray}}

\usepackage[normalem]{ulem}

\usepackage[english]{babel}

\def\beq{\begin{equation}}
\def\eeq{\end{equation}}
\def\bea{\begin{eqnarray}}
\def\eea{\end{eqnarray}}

\makeatletter
\renewcommand*{\@fnsymbol}[1]{\ensuremath{\ifcase#1\or \dagger\or *\or \ddagger\or
   \mathsection\or \mathparagraph\or \|\or **\or \dagger\dagger
   \or \ddagger\ddagger \else\@ctrerr\fi}}
\makeatother

\begin{document}

\raggedbottom
%\title{\Large The role of ribosomal proteins in determining the growth rate of bacteria and setting the bacterial cell size and shape}
%\title{\Large Interdependence of cell size, shape and bacterial growth physiology: a systems perspective}
%\title{\Large Interdependence of cell size, shape and bacterial growth physiology: proteome allocation strategies}
\title{\large Cellular resource allocation strategies for cell size and shape control in bacteria}
%\title{\Large Cell size and shape control in bacteria:\\ connections with growth physiology}
%\title{\Large Bacterial size and shape control: connections with cell growth and proteome composition}
\author{Diana Serbanescu}
\affiliation{Department of Physics and Astronomy, University College London, London WC1E 6BT, UK}

\author{Nikola Ojkic}
\affiliation{Department of Physics and Astronomy, University College London, London WC1E 6BT, UK}

\author{Shiladitya Banerjee}
\altaffiliation{Correspondence: shiladtb@andrew.cmu.edu}
\affiliation{Department of Physics, Carnegie Mellon University, Pittsburgh, PA 15213, USA}

%\linenumbers
\raggedbottom

%\noindent{\bf Abstract} \\
\begin{abstract}
\noindent Bacteria are highly adaptive microorganisms that thrive in a wide range of growth conditions via changes in cell morphologies and macromolecular composition. How bacterial morphologies are regulated in diverse environmental conditions is a longstanding question. Regulation of cell size and shape implies control mechanisms that couple the growth and division of bacteria to their cellular environment and macromolecular composition. In the past decade, simple quantitative laws have emerged that connect cell growth to proteomic composition and the nutrient availability. However, the relationships between cell size, shape and growth physiology remain challenging to disentangle and unifying models are lacking. In this review, we focus on regulatory models of cell size control that reveal the connections between bacterial cell morphology and growth physiology. In particular, we discuss how changes in nutrient conditions and translational perturbations regulate the cell size, growth rate and proteome composition. Integrating quantitative models with experimental data, we identify the physiological principles of bacterial size regulation, and discuss the optimization strategies of cellular resource allocation for size control.
\end{abstract}
\maketitle

\noindent{\bf Introduction} \\
Cell size is a fundamental physiological trait that is crucial for cellular growth, nutrient uptake, and environmental adaptation. Bacterial cells need to maintain appropriate sizes to optimize their fitness and regulate cell physiology \cite{young2006}. How cell size adapts to changes in environmental conditions is therefore a fundamental question in microbial physiology. It is known for over six decades that bacteria modulate cell sizes in response to changes in nutrient conditions. Studying the growth and morphologies of {\it{Salmonella enterica}} cells grown in different nutrient conditions, Schaechter {\it{et al.}} discovered the {\it nutrient growth law} -- the average cell size of a bacterial population increases exponentially with the population growth rate \cite{Schaechter1958}. Recent studies have confirmed this result for evolutionary divergent bacterial species such as {\it{Escherichia coli}}, {\it{Bacillus subtilis}} and \textit{Sinorhizobium meliloti} \cite{taheri2015,vadia2015,si2017,dai2018,sauls2019}, suggesting shared strategies for cell size control in bacteria. However, deviations from the nutrient growth law have been reported in studies perturbing cellular growth rate and translation via antibiotics~\cite{basan2015,si2017}. Thus the relationship between cell size and growth rate does not simply follow from the nutrient growth law and requires a deeper systems-level understanding of cellular growth physiology.

At the single-cell level, control of cell size emerges from a temporal coupling between cell growth and division. it has been established that individual cells achieve size homeostasis via a negative feedback between cell size at birth and inter-division times~\cite{amir2014,banerjee2017,taheri2015,wallden2016,cadart2019}. A particular manifestation of this principle is the {\it adder} model of cell size homeostasis, where individual cells achieve size homeostasis by adding a constant volume between successive division events~\cite{amir2014,campos2014,deforet2015,jun2015,taheri2015}. By virtue of this adder mechanism, bigger cells divide in less time than smaller cells, such that cells deviating from the homeostasis cell size quickly converge to the average cell size in a few generations. This model leads to a tight control of bacterial cell size, with the coefficients of variation ranging between 0.10 and 0.15~\cite{jun2018}. While the adder strategy for cell size homeostasis is followed by a wide range of bacterial species including {\it{E. coli}}, {\it{B. subtilis}}, {\it{Caulobacter crescentus}}, and {\it{Pseudomonas aeruginosa}} \cite{campos2014,deforet2015,jun2015,osella2014,taheri2015}), the adder model does not readily reveal a molecular basis for cell size control nor any connections between cell size and growth physiology.

In recent years, two distinct types of regulatory models have emerged that provide a molecular-level understanding of cell size control and the coupling between cell growth and division in bacteria. First is the {\it replication-initiation-centric} model~\cite{donachie1968,wallden2016,ho2015}, where cell size is determined by the time period of chromosome replication ($C$-period) and the interval between the end of chromosome replication and cell division ($D$-period). In this model, cell divides after a fixed time interval ($C+D$ period) since the initiation of chromosome replication. Second is the {\it division-centric} model for cell size control~\cite{deforet2015,ghusinga2016,harris2016,si2019,serbanescu2020,panlilio2021}, where cell division is triggered by accumulation of a threshold amount of division proteins~\cite{deforet2015, si2019, ghusinga2016,panlilio2021,ojkic2019,serbanescu2020} or cell envelope precursors~\cite{harris2016}. However, how the synthesis of division initiator proteins and $C+D$ period is controlled by the bacterial cells in different growth conditions is not well understood, leaving open the relationship between cell size, growth rate and division timing.

%Given the complexity of the processes involved in chromosome replication, multiple replication forks and cell division, the replication-initiation-centric models have been challenged in recent studies \cite{micali2018a,micali2018b,grilli2018,si2019}. In particular, It has been proposed that the slowest of two processes, one that sets replication initiation and the other controlling the division time, regulates cell size \cite{micali2018a}. Recently, with the aid of experimental data, the chromosome replication initiation and cellular division have been shown to be controlled independently in both {\it{E. coli}} and {\it{B. subtilis}} \cite{si2019}, pointing towards a {\it{division-centric}} mechanism for cell size control~\cite{deforet2015,ghusinga2016,harris2016,si2019,serbanescu2020,panlilio2021}. In this model, cell division is triggered by accumulation of a threshold amount of division proteins \cite{deforet2015,ghusinga2016,ojkic2019,si2019,panlilio2021} or cell envelope precursors \cite{harris2016}. However, how the synthesis of division proteins and cell envelop precursors are controlled by cell growth conditions is not well understood, leaving open the relationship between cell size, growth and division.

In this review, we present recent advances in quantitative modeling of bacterial cell size control, with a focus on delineating the coupling between cell size, shape and growth physiology. While there are many excellent reviews on quantitative studies of bacterial growth physiology~\cite{scott2011,scott2014,klumpp2014,taheri2015b,jun2018,bruggeman2020}, cell size regulation~\cite{sauls2016,willis2017,jun2018,cadart2019} and cell shape control~\cite{young2006,yang2016,harris2018,van2018}, the interdependence of cell morphology and growth physiology is an emerging area that that has not yet been reviewed. Here we fill this gap by reviewing regulatory models of cell division control in bacteria that reveal the connection between cell size and growth using the framework of proteome allocation. We begin by discussing quantitative laws of growth physiology that relate cellular growth rate to nutrient availability and translational capacity. Using the framework of coarse-grained proteome partitioning models, we discuss the cellular resource allocation strategies for bacterial growth control in different environments. We then present recently developed extensions of the proteome allocation model to incorporate proteins regulating cell division and cell surface synthesis. Using this framework, we present the resource allocation strategies for cell size and shape control that connect bacterial cell morphology with growth rate over a wide range of nutrient conditions and translational perturbations. \\ 

\noindent {\bf{Determinants of bacterial growth rate}}

\noindent The physiological state of a cell is characterized by its size, shape, macromolecular composition, and the rate of growth. Their interdependence is one of the key questions in bacterial physiology. The dependence of cellular growth rate on the extracellular nutrient concentration was first established by Jacques Monod for {\it E. coli} and {\it Mycobacterium tuberculosis}~\cite{monod1949}. During steady-state growth, the exponential growth rate of bacteria ($\kappa$) saturates with increasing with nutrient concentration ($c$):
\begin{equation} \label{eq:monod}
\kappa=\kappa_0\frac{c}{c+c^*}\;,
\end{equation}
\noindent where $\kappa_0$ is the maximum growth rate characteristic of the medium and $c^*$ is the nutrient concentration at half-maximum growth rate. To understand how cells achieve a nutrient-specific growth rate, a mechanistic link has to be established between the rates of nutrient import, energy production, protein synthesis and cell envelope biogenesis. Protein synthesis is essential for bacteria to proliferate. At the core of protein synthesis is the ribosome machinery which synthesizes new proteins through the process of translation. In recent work \cite{belliveau2021}, Belliveau {\it{et al.}} used proteomics data across a large number of growth conditions to examine the possible candidates that limit the growth rate of bacteria. It is found that bacterial cell growth is not limited by transporter expression for nutrient import, biosynthesis of cell envelope components, or by ATP synthesis. Instead, the translation machinery plays a crucial role, such that the synthesis of ribosomes is the rate limiting process for bacterial growth (Fig.~1A).\\

\noindent {\it Bacterial growth laws} -- In {\it{E. coli}} cells approximately 85\% of the RNA is folded into ribosomes~\cite{bremer1996}. Therefore the ratio of RNA to protein mass can provide an estimate for the ribosome mass fraction $\phi_R$ in a cell. Earlier works by Neinhardt, Magasanik and Harvey~\cite{neidhardt1960,harvey1973} uncovered a positive linear relationship between ribosome mass fraction and nutrient specific growth rate in moderate to fast growth conditions. Scott {\it{et al.}} \cite{scott2010} formalized this into a quantitative growth law of ribosome synthesis (Fig.~1B, 1F). When growth rate is modulated by changes in nutrient quality, mass fraction of ribosomes in {\it E. coli} increases linearly with the growth rate
\begin{equation} \label{eq:scott1}
\phi_R=\phi_R^{\rm min}+\frac{\kappa}{\kappa_t}\;,
\end{equation}
where $\phi_R^\text{min}$ is the mass fraction of inactive ribosomes and $\kappa_t$ is the translational capacity, defined as the average rate for amino acid chain elongation per ribosomes. The linear scaling between growth rate and ribosome mass fraction was also found in the eukaryotic budding yeast~\cite{metzl2017}. Eq.~\eqref{eq:scott1} is thus not unique to bacterial cells. 

When the growth rate of {\it E. coli} cells is altered by inhibiting translation (e.g. by adding ribosome-targeting antibiotics), ribosome mass fraction decreases linearly with growth rate (Fig. 1F, Fig. 3A)~\cite{scott2010}:
\begin{equation} \label{eq:scott2}
\phi_R=\phi_R^{\rm max}-\frac{\kappa}{\kappa_n}\;,
\end{equation}
where $\phi_R^\text{max}$ is the maximum mass fraction of ribosomal proteins, and $\kappa_n$ is the nutritional capacity of the medium that has a positive correlation with the nutrient-specific growth rate. The observation that the maximum mass fraction of ribosomal proteins $\phi_R^\text{max}\approx 0.55$ is much below 1, suggests a coarse-grained model of proteome partitioning into three components: ribosome-affiliated proteins (R-sector) of mass fraction $\phi_R$, housekeeping proteins (Q-sector) of mass fraction $\phi_Q$ that are not affected by translation inhibition, and the remaining non-ribosomal proteins (P-sector) whose mass fraction $\phi_P\rightarrow 0$ as $\phi_R \rightarrow \phi_R^\text{max}$ (Fig.~1C and D). Given the invariance of $\phi_Q$ with translational perturbations, the $R$- and $P$- sectors add up to a constant such that $\phi_R+\phi_P=\phi_R^{\rm max}=1-\phi_Q$. 

\begin{figure*}[htp!]
  \includegraphics[width=\linewidth]{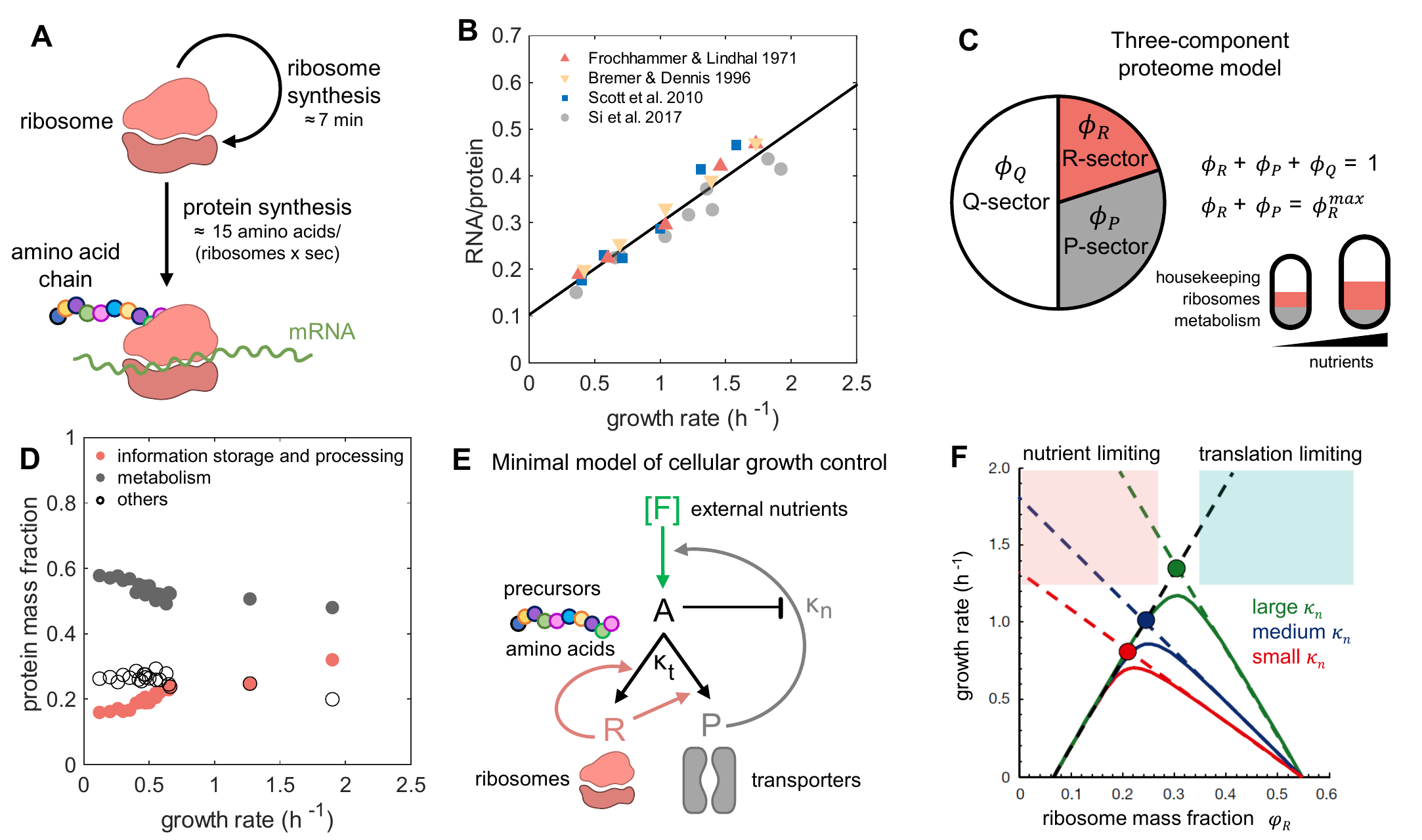}
  \vspace{-10pt}
  \caption{{ \bf Growth rate regulation in bacteria.}
{\bf{A.}} Schematic showing translation and ribosomal protein synthesis, which constitute the limiting factors for bacterial cell growth. 
{\bf{B.}} Ribosomal mass fraction in {\it E. coli}, approximated by the ratio of RNA to protein mass, increases linearly with the growth rate of the cell modulated by nutrients. Data taken from refs.~\cite{forchhammer1971,bremer1996,scott2010,si2017}.
{\bf{C.}} The three-component proteome partitioning model \cite{scott2010}, consisting of the Q-sector that is invariant under translational perturbations, the R-sector of ribosomal proteins that increase in mass fraction with increasing growth rate, and the P-sector of non-ribosomal proteins that is constrained by the relation: $\phi_P+\phi_R=\phi_R^\text{max}=1-\phi_Q\approx 0.55$. 
{\bf{D.}}  Proteomic data \cite{schmidt2016} 
show a decrease in the mass fraction of metabolic sector proteins with growth rate, at the cost of an increase in mass fraction of ribosomal sector proteins (information storage and processing), consistent with the three-component proteome allocation model. 
{\bf{E.}} Schematic of a minimal model for cellular growth control. 
Precursor molecules (amino acids, A) are produced by the action of transporters (P) on the external nutrients. Transporters (P) and ribosomal proteins (R) are synthesized from the precursors by the ribosomes R.
{\bf{F.}} The optimal value of the ribosomal protein mass fraction $\phi_R$ depends on the growth environment (poor nutrient - red line, moderate nutrients - blue, rich nutrient - green). Dashed lines correspond to the equation: $\kappa=\kappa_t(\phi_R-\phi_R^\text{min})$ and $\kappa=\kappa_n (\phi_R^\text{max}-\phi_R)$. For each nutrient condition the growth rate exhibits a maximum (solid circle) corresponding to an optimal allocation of $\phi_R$. The upper bound on the growth rate maximization occurs when the translation efficiency $\kappa_t$ and the nutritional efficiency $\kappa_n$ are both maximum for a given nutrient environment. Figure adapted from Ref.~\cite{scott2014}.}
  \label{Figure:1}
\end{figure*}

During steady-state exponential growth, the rate of amino acid supply by the P-sector proteins must be balanced by the rate of amino acid consumption by the R-sector proteins, such that the amino acid pool is maintained at constant size. If the amino acid pool increases such that the supply exceeds demand, then $\phi_R$ increases to meet the demand of protein biosynthesis and simultaneously decrease supply due to the constraint $\phi_P=\phi_R^{\rm max}-\phi_R$ \cite{scott2014}. This strategy underlies flux balance and maximization of the growth rate, as discussed in the next section. The partitioning of the proteome into three main components is supported by mass spectrometry data \cite{schmidt2016} (Fig.~1D) where clustering the proteins based on their function reveals similar behaviour to the proposed model components. More precisely, the category identified as ``information storage and processing'', which contains ribosome-affiliated proteins, increases with growth rate, while the ``metabolism'' cluster decreases with growth rate, equivalent to the non-ribosomal proteins in the $P$-sector. This leaves the mass fraction of the rest of the proteins to be independent of growth rate, akin to the housekeeping $Q$-sector. \\

\noindent {\bf{Proteome allocation strategies for cellular growth control}}\\
The proteome partitioning model~\cite{scott2010} together with the empirical growth laws (Eq.~\ref{eq:scott1} and \ref{eq:scott2}) generates constraints on cellular resource strategies linking cell growth, nutrient uptake, metabolism and protein synthesis. This framework provides a theoretical basis for explaining the empirical growth laws (Eqs. 2 and 3). The positive linear relationship between growth rate and ribosome mass fraction follows from exponential growth of total protein mass $M$ at steady-state: ${\rm d}M/{\rm d}t=\kappa M$. The rate of accumulation of protein mass is balanced by the rate of protein synthesis by active ribosomes $\kappa M=\kappa_t (M_R - M_R^\text{min})$, where $M_R^\text{min}$ is the mass of inactive ribosomes, leading to the first empirical growth law in Eq.~\eqref{eq:scott1}. The second growth law (Eq. \ref{eq:scott2}) follows from the constraint on amino acid flux to meet the demand of protein synthesis during exponential growth~\cite{molenaar2009,scott2010,scott2014} (Fig.~1E). The rate of change of amino acid concentration ($A$) in a cell is given by:
\begin{equation}\label{eq:A}
\frac{{\rm d}A}{{\rm d}t}=\kappa_n (A) \phi_P-\kappa_t (A) (\phi_R-\phi_R^\text{min})\;,
\end{equation}
where $\phi_P=\phi_R^\text{max}-\phi_R$ and both the translational capacity ($\kappa_t$) and the nutritional capacity ($\kappa_n$) are functions of the amino acid concentration $A$. In particular, $\kappa_t$ increases and then saturates with $A$~\cite{klumpp2013,scott2014}, whereas there is a negative feedback control of $\kappa_n$ to limit the concentration of amino acids~\cite{neidhardt1990}. At steady-state, there is no net accumulation of amino acids in the cell: the nutrient import needs to meet the demands of synthesizing new proteins \cite{scott2010,scott2014,ni2020}, resulting in the growth law: $\kappa=\kappa_t (\phi_R-\phi_R^\text{min})=\kappa_n (\phi_R^\text{max}-\phi_R)$. The flux balance condition then determines the relationship between steady-state ribosome mass fraction and amino acid concentration:
\begin{equation}
\phi_R=\frac{\kappa_n(A) \phi_R^\text{max} + \kappa_t(A)\phi_R^\text{min}}{\kappa_n + \kappa_t}\;.
\end{equation}
As a result the steady-state value of $\phi_R$ decreases with $A$, but there is no unique value of ribosome mass fraction that satisfies the amino acid flux in a given growth condition. How do then bacterial cells regulate the ribosomal mass fraction $\phi_R$? The optimal value of ribosome mass fraction is set by the maximum achievable growth rate~\cite{molenaar2009,scott2014}, which exhibits a unique maximum as a function of $\phi_R$ In a given growth environment (Fig.~1F). This growth rate maxima is reached when both the translational and nutrient capacities are at their maximal value, and increases with increasing $\kappa_n$.
%For each growth condition there is a pair $(\kappa_n,\kappa_t)$ that maximizes the growth rate \cite{scott2014} (Fig.~1F). Therefore, for each ribosome protein fraction $\phi_R$, there is a maximum achievable growth rate and this value is given by Eq.~\eqref{eq:scott1}. 
In similar spirit, mechanistic models of growth rate optimization and ribosome regulation have been developed by others~\cite{bosdriesz2015,maitra2015,weisse2015,kohanim2018}.
%\begin{equation}
%\kappa_n (a) \phi_P=\kappa_t (a) (\phi_R-\phi_R^\text{min})=\kappa
%\end{equation}
%where $\kappa_n$ is the nutritional capacity of the cell, $\kappa_t$ is the translation capacity, $\phi_R^\text{min}$ is the mass fraction of inactive ribosomal proteins and $\phi_P=\phi_R^\text{max}-\phi_R$. Both $\kappa_n$ and $\kappa_t$ depend on the amino acid concentration ($a$) inside the cell, such that $\kappa_n(a)=\kappa_n^0f(a)$, $\kappa_t=\kappa_t^0 g(a)$, where $f(a)$ and $g(a)$ are regulatory functions with values between 0 and 1, $\kappa_t^0$ and $\kappa_n^0$ are the maximum values of $\kappa_t$ and $\kappa_n$ that depend on the growth medium.

Taking further the idea of compartmentalizing the proteome, Pandey and Jain \cite{pandey2016} developed a {\it precursor-transporter-ribosome} model, where they considered the coupled dynamics of transporters ($P$) that import nutrients from the extracellular medium and convert them into amino acid precursors ($A$). The amino acids are then converted into transporters and ribosomal proteins ($R$) (Fig.~1E). The ribosomes in turn catalyze the production of both transporters and ribosomes, while the transporters catalyze the production of more precursors \cite{pandey2016}. The coupled dynamics of transporters (mass $M_P$) and ribosomal proteins (mass $M_R$) are given by
%\begin{equation} \label{eq:pandeyP}
%\frac{{\rm d}P}{{\rm d}t}=k_P T-\kappa \frac{RP}{V}
%\end{equation}
\begin{equation} \label{eq:pandeyT}
\frac{{\rm d}M_P}{{\rm d}t}=\kappa_t f_P(M_R - M_R^\text{min}) -d_P M_P
\end{equation}
\begin{equation} \label{eq:pandeyR}
\frac{{\rm d}M_R}{{\rm d}t}=\kappa_t f_R(M_R - M_R^\text{min})-d_R M_R
\end{equation}
where $d_P$ and $d_R$ are the degradation rates for the transporters and the ribosomes, respectively, $f_R$ is the fraction of ribosomes engaged in the production of ribosomal proteins, and $f_P$ is the fraction of ribosomes catalyzing the production of transporters. The parameters $f_R$ and $f_P$ are subjected to the constraint: $f_P+f_R=\phi_R^\text{max}$, where the choice of $f_R$ can be determined by the regulatory condition of growth rate maximization~\cite{molenaar2009}. In a given growth environment with other cellular parameters fixed, the regulation adjusts the value of $f_R$ such that the growth rate is maximized (Fig.~1F). The optimized steady-state of the precursor-transporter-ribosome model then reproduces the empirical growth laws (Eq.~\ref{eq:scott1} and \ref{eq:scott2})~\cite{pandey2016}. Thus, the proteome allocation theory provides a promising framework to understand the relationships between the growth rate and macromolecular composition of the bacterial cell.\\

\noindent {\bf{Interdependence of cell size and growth rate}}

\noindent The relationship between ribosome abundance and growth rate does not readily offer any information about cell size regulation. To understand the control of cell size in different growth environments, we present a model linking cellular growth and protein synthesis to cell division control. Growth of single bacterial cells have been observed to be exponential in many species including {\it E. coli}, {\it B. subtilis} and {\it C. crescentus}~\cite{taheri2015,wright2015,wang2010}. Linear growth of cell size is usually encountered in mammalian cells~\cite{cadart2018} and is not considered here. We therefore model an exponentially growing cell, which elongates exponentially in length during its cell cycle while maintaining a constant width (Fig. 2A). The volume ($V$) of the cell grows exponentially as
\begin{equation} \label{eq:V1}
\frac{{\rm d}V}{{\rm d}t}=\kappa(\phi_R) V(t)
\end{equation}
where the growth rate $\kappa$ is a function of the ribosome mass fraction $\phi_R$, as defined in Eq.~\eqref{eq:scott1}. The cell volume increments by an amount $\Delta V=V(0)(e^{\kappa t}-1)$ between cell birth ($t=0$) and division ($t=\tau$). Single-cell studies on {\it E. coli}, {\it B. subtilis}, {\it P. aeruginosa} and many other bacterial species~\cite{willis2017} have revealed that bacterial cells divide after adding a fixed volume $\Delta V$ between consecutive division events, irrespective of cell size at birth \cite{amir2014,deforet2015,banerjee2017,taheri2015,wallden2016}. Regulatory models for this division control mechanism have recently been proposed~\cite{deforet2015,basan2015,ghusinga2016}, where cell size is regulated by the abundance of a putative division protein. The number of division proteins grows in proportion to the cell size and the cell divides once a threshold amount of division proteins has been accumulated (Fig.~2A). In this {\it threshold initiation model}, the dynamics of the division protein copy number $X$ is described the following equation:
\begin{figure*}[htp!]
  \includegraphics[width=\linewidth]{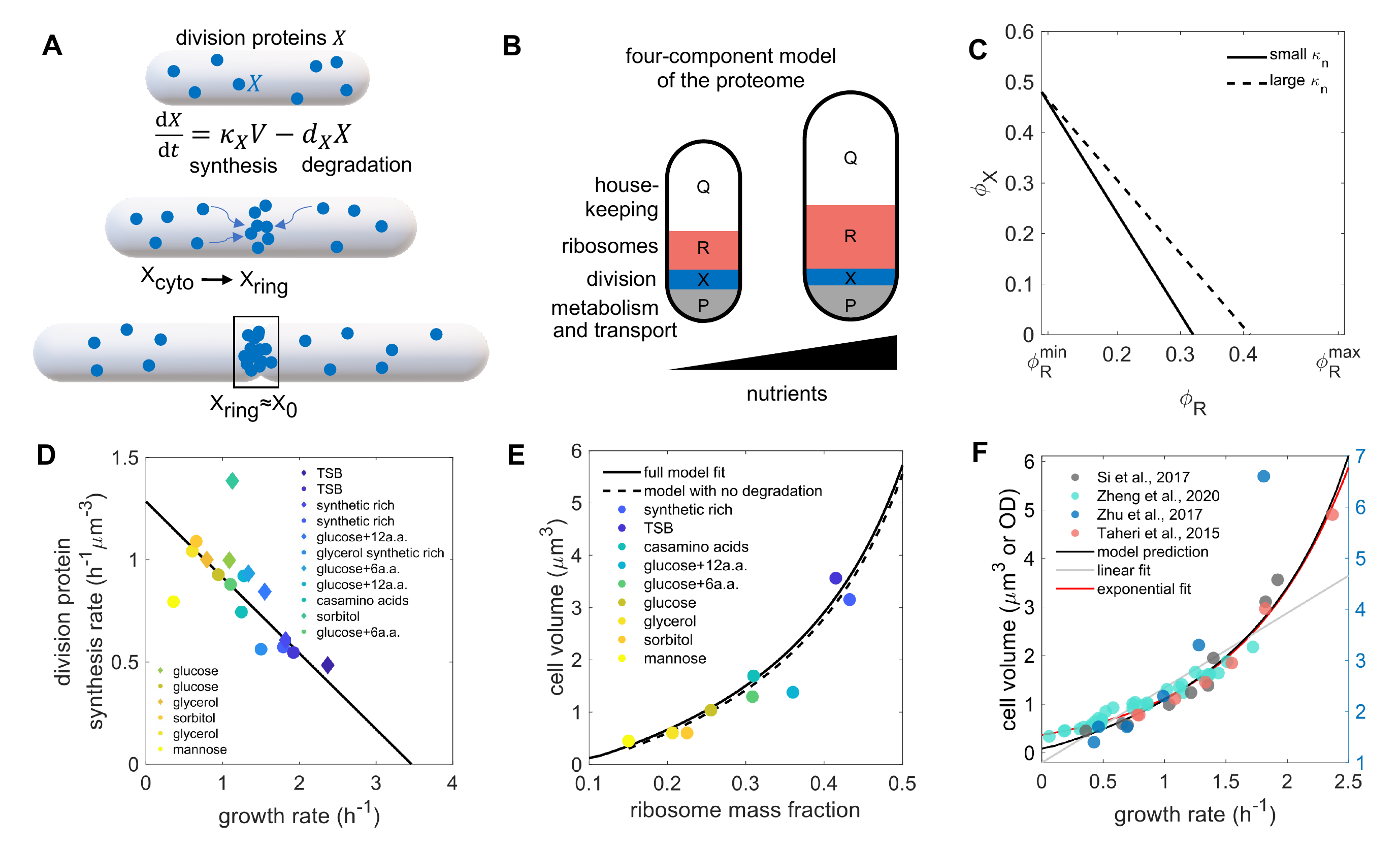}
  \vspace{-10pt}
  \caption{{ \bf Interdependence of cell size and growth rate under nutrient modulations.}
{\bf{A.}} Schematic of the {\it{threshold initiation}} model of cell division, where division proteins $X$ are produced in the cytoplasm at a rate proportional to cell volume, and recruited at the middle of the cell to form the division ring. Once a threshold amount of $X$ proteins is accumulated, the cell divides. 
{\bf{B.}} Schematic of the four-component proteome partitioning model, where a new proteome sector for division proteins is included~\cite{scott2010}. The division sector size decreases with nutrient quality as cells promote the production of more ribosomes.
{\bf{C.}} Decrease in division protein mass fraction with increasing ribosomal protein mass fraction for small and large nutritional capacities.
{\bf{D.}} Linear decrease in division protein synthesis rate with the nutrient-imposed growth rate suggesting a tradeoff between resources allocated for division and growth.
{\bf{E.}} Fitted model for average cell volume as a function of ribosome mass fraction. Data taken from ref.~\cite{si2017}. Solid line: full model fit with $X$ protein degradation; dashed line: approximate model with no $X$ degradation \cite{serbanescu2020}.
{\bf{F.}} Comparison between the model prediction \cite{serbanescu2020} for cell volume as a function of growth rate and the best fit linear \cite{zheng2020} and exponential \cite{Schaechter1958} functions. Data are taken from Refs. \cite{si2017, zheng2020, zhu2017, taheri2015}
}
  \label{Figure:2}
\end{figure*}
\begin{equation} \label{eq:X}
\frac{{\rm d}X}{{\rm d}t}=k_X(\phi_R)V(t)-d_X X(t)
\end{equation}
\noindent where $k_X$ is the volume specific rate of synthesis of the division proteins and $d_X$ is their degradation rate. Cell divides when $X$ reaches a fixed threshold amount $X_0$. In the limit when $\kappa\gg d_X$, $\Delta V=X_0\kappa/k_X$.  As $k_X$, $\kappa$ and $X_0$ are constant for a given growth medium, cells add a constant volume $\Delta V$ in each growth generation, consistent with the adder model. In the opposite limit when $\kappa\ll d_X$, cell size at division $\approx X_0d_X/k_X$, consistent with data that {\it E. coli} behaves sizer-like in slow growing media~\cite{wallden2016,si2019}. While the threshold initiation model describes adder and sizer behaviors it does not capture mixed growth modes observed in some bacteria, such as in {\it C. crescentus}, where the adder mechanism is only implemented for part of the cell cycle~\cite{banerjee2017}.

Based on the threshold initiation model, the average cell volume $V$ in a given growth medium is given by
\begin{equation}\label{eq:V}
V=\frac{\kappa+d_X}{\kappa_X(2-2^{-d_X/\kappa})}\;,
\end{equation}
where we define $\kappa_X=k_X/(2X_0\ln{2})$ as the normalized rate of division protein synthesis. Eq.~\eqref{eq:V} underlies the interdependence of cell volume and growth rate, where cell volume can be perturbed by modulating both $\kappa$ and $\kappa_X$. The above expression for the average cell volume is not specific to a particular organism but is applicable to all exponentially growing bacterial cells that achieve cell size homeostasis via the adder principle or threshold accumulation of division proteins. We therefore ask next how the rate of division protein synthesis is controlled by cellular growth rate.\\

\noindent{\bf Proteome allocation strategies for cell size control}\\
Proteome allocation theory provides a useful framework to understand how the main protein components of the cell's proteome vary with growth rate. To uncover the relationship between cell division control and growth rate, we present a recently developed extended proteome sector model~\cite{basan2015,serbanescu2020,bertaux2020} that couples the rate of division protein synthesis to ribosomal translation. In the extended proteome sector model (Fig.~2B), a separate sector is included for the division proteins, subject to the constraint that the sum of sectors' mass fractions add to 1: $\phi_P+\phi_R+\phi_Q+\phi_X=1$. The total mass of the division proteins, $M_X$, grows at a rate proportional to the mass of actively translating ribosomes:
\begin{equation} \label{eq:mx}
\frac{{\rm d}M_X}{{\rm d}t}=\kappa_t f_X(M_R - M_R^\text{min})-d_X M_X\;,
\end{equation}
where $f_X$ is the fraction of ribosomes synthesizing division proteins that equals $\phi_X$ at steady-state. Following Serbanescu \textit{et al.}~\cite{serbanescu2020}, the mass fraction of $X$ at steady-state can be determined using the
 %that equals $\phi_X$ at steady-state. Comapring Eq.~\eqref{eq:mx} with Eq.~\eqref{eq:X} 
 %we can identify that the rate of division protein synthesis is proportional to the mass fraction of the division protein sector: $\kappa_X \propto \phi_X \kappa_t (\phi_R-\phi_R^\text{min})$. 
flux balance equation %(Eq.~\eqref{eq:A}) 
\begin{equation}
\kappa_n(\phi_R^\text{max}-\phi_R-\phi_X)=\kappa_t(\phi_R-\phi_R^\text{min})\;.
\end{equation}
\noindent This leads to the following relation between $\phi_X$ and ribosome mass fraction:
\begin{equation}\label{eq:fx}
\phi_X=\frac{\kappa_n \phi_R^\text{max}+\kappa_t \phi_R^\text{min}}{\kappa_n}-\frac{\kappa_n+\kappa_t}{\kappa_n}\phi_R
\end{equation}
predicting a negative correlation between $\phi_X$ and $\phi_R$ under nutrient perturbations (Fig.~2C), consistent with recently published proteomics data~\cite{mori2021}. In an alternative approach, Bertaux \textit{et al.}~\cite{bertaux2020} assumed that the mass fraction of the division sector depends on the concentration of two other proteome sectors, since the dependency of division protein concentration on a single proteome sector was inadequate to capture the behaviour of cell size under growth perturbations. Therefore, the authors assumed a
phenomenological functional form $\phi_X=\phi_P^\alpha\phi_R^\beta$, where the parameters $\alpha$ and $\beta$ are deduced by fitting data. Both approaches lead to qualitatively similar results for the dependency of cell division sector on growth rate under nutrient perturbations.

Combining Eq.~\eqref{eq:X}, \eqref{eq:mx} and \eqref{eq:fx} the following relation emerges for the dependence of division protein synthesis rate on the mass fraction of division proteins~\cite{serbanescu2020}:
\begin{equation} \label{eq:kp}
\kappa_X=\kappa_X^0(\phi_R^\text{M}-\phi_R)
\end{equation}
\noindent where $\kappa_X^0$ is the rate of production of division proteins per ribosomes, and $\phi_R^\text{M}=\frac{\kappa_n \phi_R^\text{max}+\kappa_t \phi_R^\text{min}}{\kappa_n+\kappa_t}$. By combining Eqs.~\eqref{eq:scott1} and \eqref{eq:kp} we obtain a central result that there is a negative correlation between the rate of synthesis of division proteins and growth rate of the cell (Fig.~2D), consistent with experimental data on \textit{E. coli} growth physiology \cite{si2017}. This suggests a nutrient-dependent tradeoff between the rates of two main physiological processes in the cell growth and division. This principle underlies the control of cell size, which can be expressed as a function of ribosome mass fraction 
\begin{equation}
V=\frac{\kappa_t (\phi_R-\phi_R^\text{min})+d_X}{\kappa_X^0(\phi_R^\text{M}-\phi_R)(2-2^{-d_X/\kappa_t(\phi_R-\phi_R^\text{min})})}
\end{equation}
such that average cell volume increases with increasing ribosome mass fraction (Fig.~2E, solid line). Fitting this model with experimental data on \textit{E. coli}~\cite{si2019} one gets $d_X=0.24$h$^{-1}$, allowing for the approximation $V\approx \kappa/\kappa_X$ in moderate to fast growing bacteria (Fig.~2E, dashed line). We can then express cell volume as a function of growth rate, recapitulating the celebrated nutrient growth law~\cite{Schaechter1958} that the cell size increases nonlinearly with growth rate (Fig.~2F): 
\begin{equation}
V\approx\frac{\kappa/\kappa_X^0}{(\phi_R^\text{M}-\phi_R^\text{min})-\kappa/\kappa_t}\;.
\end{equation}

This result captures the departure from an exponential relationship between cell size and growth rate~\cite{Schaechter1958} in slow growing medium, where a linear relationship emerges. However, a linear relationship between cell size and growth rate, as recently proposed by Zheng et al~\cite{zheng2020}, does not accurately capture the cell size data for fast growing bacterial cells \cite{zhu2017}.\\
%Proteomic models are extremely useful in explaining how the main proteome sectors vary with growth rate. Combined with the {\it{adder}} or {\it{initiator}} models for division control, they are a useful tool in understanding how cell size regulation arises under nutrient perturbations. 

\noindent {\bf Molecular basis for cell division and size control}\\
%Following a division-centric model, the adder model can be explained at molecular level by a threshold accumulation of cell envelope precursors \cite{harris2016} or division proteins such as FtsZ \cite{ojkic2019,si2019} (Fig.~2A).
\textit{FtsZ as cell size sensor} -- The threshold initiation model for cell size control is agnostic about the identity of the division protein $X$. While various proteins could be potential candidates for cell division initiation~\cite{adams2009,bi1991,flaatten2015}, a recent study has identified the protein FtsZ as the key size sensor molecule that reaches a threshold abundance at the time of cell division in \textit{E. coli} \cite{si2019}. FtsZ assembles a ring-like structure in the mid-cell, triggering septation at a critical size of the ring.
%division initiator that assembles a ring-like structure in the mid-cell region to trigger septation~\cite{si2019}. 
FtsZ accumulation in the cell can be regarded as a two-stage process: accumulation of FtsZ in the cytoplasm, with abundance $X_\text{cyto}$, followed by recruitment at the middle of the cell, with abundance $X_\text{ring}$ (Fig.~2A). The dynamics of the cytoplasmic and ring-bound FtsZ are given by \cite{si2019,ojkic2019,ojkic2021}:
\begin{equation}
\frac{{\rm d}X_\text{cyto}}{{\rm d}t}=-k_b X_\text{cyto}+k_d X_\text{ring}+\kappa_X V -d_X X_\text{cyto}
\end{equation}
\begin{equation}
\frac{{\rm d}X_\text{ring}}{{\rm d}t}=k_b X_\text{cyto}-k_d X_\text{ring} -d_X X_\text{ring}
\end{equation}
\noindent where $k_b$ is the binding rate of cytoplasmic FtsZ to the Z-ring, $k_d$ is the disassembly rate and $\kappa_X$ is the rate of synthesis of FtsZ. Defining $X=X_\text{cyto}+X_\text{ring}$, one recovers the dynamics for division protein synthesis in Eq.~\eqref{eq:X}, with the underlying assumption that the timescale for FtsZ recruitment to the division ring is much faster than the cytoplasmic production. In \textit{E. coli}  cells, it has been shown that cell division is triggered once a fixed threshold amount of proteins is accumulated in the division ring, $X_\text{ring}=X_0$ \cite{si2019}, with $X_0$ scaling with the width of the cell (Fig.~2A) \cite{ojkic2019}.
%Cell division is triggered once a fixed threshold amount of proteins is accumulated in the division ring, $X_\text{ring}=X_0$, with $X_0$ scaling with the width of the cell (Fig.~2A)~\cite{ojkic2019}. 
Thus the added cell length in each division cycle is coupled to the cell width, resulting in conservation of cellular aspect ratio~\cite{ojkic2019}. %We return to this point later.

Direct quantification of the rate of division protein production in different growth media is currently lacking. The division protein synthesis rate $\kappa_X$ can be measured directly by quantifying the rate of accumulation of FtsZ in the cytoplasm. This can be done by fluorescently labeling FtsZ~\cite{si2019} and measuring the rate of change in intensity per volume in different growth conditions. Recently Panlilio \textit{et al.}~\cite{panlilio2021} demonstrated that threshold accumulation of a constitutively expressed P-sector division protein triggers cell division. In this case, $\kappa_X$ has been estimated by measuring the production rate of GFP in strains having chromosomal reporters for constitutive promoters. \\

\noindent\textit{Cell size control by metabolic sensors} -- While the proteome allocation model in combination with the threshold initiation model quantitatively predicts cell size regulation in \textit{E. coli} in different nutrient environments, these models cannot be simply extended to other bacterial organisms such as the Gram-positive \textit{B. subtilis}. Weart \textit{et al.}~\cite{weart2007} 
discovered that in \textit{B. subtilis}, FtsZ accumulation in the Z-ring is controlled by the protein UgtP that acts as a FtsZ inhibitor in a nutrient-dependent manner. UgtP is a metabolic sensor whose affinity for FtsZ is sensitive to the available nucleotide sugar (UDP-glucose). In nutrient-poor conditions, UgtP mostly binds to itself forming homodimers and weakly affects the FtsZ pool. While in nutrient-rich conditions UgtP strongly inhibits FtsZ accumulation in the ring, delays cell division resulting in elongated cells~\cite{chien2012}. 
Recently, Ojkic and Banerjee~\cite{ojkic2021} developed a molecular model for cell size control in \textit{B. subtilis} by FtsZ inhibitors, predicting that UgtP synthesis rate increases non-linearly with cellular growth rate, in line with recent experimental findings~\cite{hill2018}. 
This nonlinear dependence of UgtP production rate on growth rate leads to an increase in cell length and aspect ratio with nutrient concentrations, as observed experimentally~\cite{sauls2019,sharpe1998}.\\

\noindent\textit{Cell size control by chromosome replication} -- In the \textit{replication-initiation-centric} models of cell size control in {\it E. coli} \cite{wallden2016, donachie1968, ho2015,cooper1968}, 
cell size is determined by the time period of chromosome replication ($C$-period) and the time period from replication termination to cell division ($D$-period). Cell division is triggered after a fixed time ($C+D$ period) has elapsed since the initiation of chromosome replication. As a result, cell size at division is given by $V_i e^{\kappa(C+D)}$ , where $\kappa$ is the growth rate, and $V_i$ is the cell size at the initiation. If the $C+D$ period and the cell size at initiation is invariant under different nutrient conditions~\cite{cooper1968, cooper1997}, 
then the exponential relationship~\cite{Schaechter1958} between cell size and growth rate is recovered. A constant $C+D$ period however indicates a sizer model where cells grow to a fixed size independent of cell size at birth, inconsistent with the adder model. When the constraint of a fixed initiation size is replaced by an adder principle for the control of cell mass at the initiation of DNA replication~\cite{ho2015}, then the adder model for cell division control can be recovered. The invariance of $C+D$ period with nutrient conditions has recently been questioned in two different studies~\cite{wallden2016,zheng2020}, with data suggesting that $C+D$ period is inversely proportional to the growth rate.

Given the complexity of the processes involved in chromosome replication, multiple replication forks and cell division, the replication-initiation-centric models have been challenged in recent studies~\cite{si2019, micali2018a, micali2018b, grilli2018, witz2019}. 
In particular, it has been proposed that the slowest of two processes, one that sets replication initiation and the other controlling the division time, regulates cell size~\cite{micali2018b}. 
Recently, a \textit{replication double adder} model has been proposed~\cite{witz2019}, 
where cells grow by a fixed volume per replication origin between two consecutive initiation cycles, and cell divides after elongation by a constant volume per origin of replication. The replication double adder model, however, is inconsistent with the adder principle of cell size homeostasis and predicts a more sizer-like behavior~\cite{le2021}. 
With the aid of experimental data in varying growth conditions, the chromosome replication initiation and cellular division have been shown to be controlled independently in both \textit{E. coli} and \textit{B. subtilis} \cite{si2019}, pointing towards an \textit{independent double adder} model~\cite{micali2018b}. 

The connection between $C+D$ period and division protein synthesis was revealed in a recent model proposed by Zheng \textit{et al.}~\cite{zheng2020}, where the mass added between birth and division $\Delta m$ is given by $\Delta m=m_0 (C+D)$, with $m_0$ a constant. This equation predicts that the added cell volume $\Delta V$ is proportional to $\kappa (C+D)$. For this model to be consistent with the division sector model~\cite{serbanescu2020}, we expect $C+D\propto(\phi_R^M-\phi_R )^{-1}\propto \kappa_X$, which provides a direct link between ribosomes, division protein synthesis rate and the chromosome replication period.\\

\noindent {\bf{Cell size control under translational perturbations}}

\noindent Translation inhibition is one of the most common modes of antibiotic action~\cite{mccoy2011}. When bacteria are subjected to translation inhibitory antibiotics, cells grow at a reduced rate while undergoing changes in cell size and shapes~\cite{si2017,basan2015,harris2016,scott2010,banerjee2021}. For instance, Gram-negative \textit{E. coli}, \textit{C. crescentus} and the Gram-positive \textit{L. monocytogenes} decrease their surface-to-volume ratio upon exposure to translation inhibitory antibiotics~\cite{si2017,harris2016}. Understanding the coupling between cell morphology, growth and translation presents new challenges in bacterial growth physiology.
%\noindent{\bf{Translational control of cell size}}
%
%\noindent Protein synthesis is essential for bacteria to proliferate. At the core of protein synthesis is the ribosome machinery which synthesizes new proteins through the process of translation. Inhibiting translation is therefore one of the most common modes of antibiotic action~\cite{mccoy2011}. When bacteria are subjected to translation inhibitory antibiotics, cells grow at a reduced rate while undergoing changes in cell size and shapes~\cite{scott2010,basan2015,harris2016,si2017,banerjee2021}. Understanding the coupling between cell morphology, growth and translation presents new challenges in bacterial growth physiology.
%the protein synthesis is inhibited and therefore a change in the cell's morphology is expected as a result. Even though translation inhibitory antibiotics are widely used, their effects on bacterial size and shape lack a quantitative description. 

\begin{figure*}[htp!]
  \includegraphics[width=\linewidth]{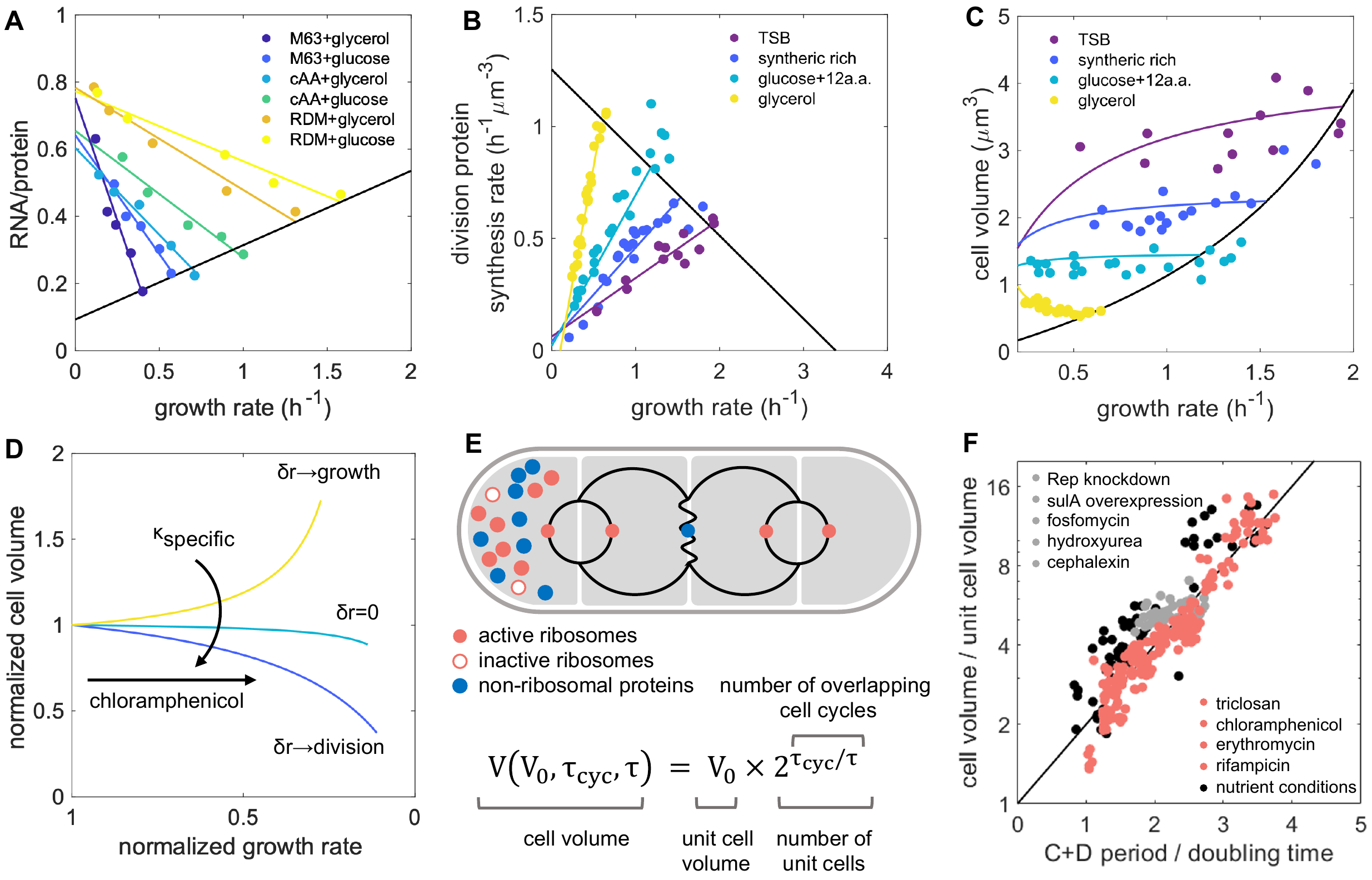}
  \vspace{-10pt}
  \caption{{ \bf Cell size control under translation inhibition.}
{\bf{A.}} Under translation inhibition, cellular growth rate decreases while the ribosome mass fraction increases. Data taken from Scott et al~\cite{scott2010}. 
{\bf{B.}} Model (Eq.~\ref{eq:kp2}) fit to experimental data~\cite{si2017} for division protein synthesis rate as a function of growth rate in four different nutrient conditions under translation inhibition \cite{serbanescu2020}.
{\bf{C.}}Volume as a function of growth rate under translation inhibition \cite{serbanescu2020}.
{\bf{D.}} Three distinct morphological responses to translation inhibition by chloramphenicol, depending on the quality of nutrients. Volume and growth rates are normalized by their initial values before chloramphenicol is applied. In nutrient-rich media, cells allocate more ribosomes to division (dark blue line), thus increasing the surface-to-volume ratio to promote nutrient influx, whereas in nutrient-poor media, they allocate more ribosomes towards growth, inflating the cell size (yellow line) and in turn decreasing the surface to volume ratio to reduce antibiotic influx \cite{serbanescu2020}.
{\bf{E.}} Schematic adapted from Si \textit{et al.} \cite{si2017} showing the general growth law, where cell volume $V$ is related to the volume $V_0$ of unit cells as: $V(V_0,\tau_{cyc},\tau)=V_0 2^{\tau_{cyc}/\tau}$, where $\tau_{cyc}$ is the cell cycle duration ($C+D$ period), and $\tau$ is the cell doubling time. Each unit cell contains sufficient resources for self-replication (active ribosomes, inactive ribosomes and non-ribosomal proteins). The number of unit cells correlates with the number of overlapping cell cycles such that fast growing cells that initiate multiple rounds of DNA replication have more unit cells. 
{\bf{F.}} Experimental validation of the general growth law. After rescaling cell volume by the unit cell volume $V_0$ and the growth rate by $\tau_{cyc}^{-1}$, all data obtained for translation perturbations and other types of perturbation (see legend) collapse onto the master curve $V/V_0= 2^{\tau_{cyc}/\tau}$.} 
  \label{Figure:3}
\end{figure*}

Quantitative studies in recent years have made great progress in defining the relationship between cellular growth rate and ribosome concentration under translation inhibition by ribosome-targeting antibiotics~\cite{scott2010,elf2006,greulich2015,greulich2017}. To compensate for the ribosomes bound by antibiotics, bacteria produce more ribosomes to increase the abundance of ribosome-affiliated proteins in the R-sector. As defined in Eq.~\eqref{eq:scott2}, experimental data on \textit{E. coli} reveal a linear relationship between ribosome mass fraction ($\approx$RNA/Protein) and growth rate~\cite{scott2010} (Fig.~3A).
%(Fig.~3A):
%\begin{equation} \label{eq:scott2}
%r=r_{max}-\frac{\kappa}{\kappa_n}
%\end{equation}
%\noindent where $r_{max}$ is the vertical intercept and $\kappa_n$ is the nutritional capacity and exhibits a strong positive correlation to the growth rate. $r_{max}$ can be interpreted as the limit of the ribosomal protein sector when the translation capacity is reduced to zero (i.e. cells do not grow). 
In the presence of a division protein sector, the relationship between $\kappa$ and $\phi_R$ becomes
\begin{equation}
\kappa=\kappa_n(\phi_R^\text{max}-\phi_X-\phi_R)\;,
\end{equation}
where the nutritional capacity $\kappa_n$ increases with the growth rate imposed by the medium \cite{scott2010}. It follows that the division protein synthesis rate is positively correlated to the cellular growth rate under translation inhibition~\cite{serbanescu2020} (Fig.~3B):
\begin{equation} \label{eq:kp2}
\kappa_X=\kappa_X^0\left(\frac{\kappa}{\kappa_n}+\delta r\right)
\end{equation}
\noindent where $\delta r=\phi_R^\text{M}-\phi_R^\text{max}+\phi_X$ is interpreted as the excess ribosomal mass fraction allocated to division protein synthesis under translation limitation. Translation inhibition disrupts the balanced allocation of resources and tradeoff between growth and division. Depending on whether the excess ribosomes ($\delta r$) are allocated to growth or division, an increase or a decrease in cell size is observed (Fig.~3C). Using proteome allocation theory presented earlier, it can be derived that $\delta r$ increases with the nutrient specific growth rate, such $\delta r>0$ in nutrient rich medium and $\delta r<0$ in nutrient poor medium~\cite{serbanescu2020}. Combining Eq.~\eqref{eq:kp2} with Eq.~\eqref{eq:V} one can then predict cell volume changes under translation inhibition
\begin{equation}
V\approx \frac{\kappa/\kappa_X^0}{\kappa/\kappa_n +\delta r}\;.
\end{equation}
The above expression for cell volume suggests three distinct morphological responses to translation inhibition \cite{serbanescu2020} (Fig.~3D). In nutrient-poor media (Fig.~3D, yellow), more ribosomes are allocated towards growth than division ($\delta r<0$), resulting in increased cell size upon growth inhibition. In intermediate nutrient concentrations, resources are balanced between growth and division leading to an invariance in cell size. The latter has been reported in Basan \textit{et al.}~\cite{basan2015} for {\it E. coli} under chloramphenicol treatment. In nutrient-rich media (Fig.~3D, dark blue), more ribosomes are allocated towards division ($\delta r>0$), resulting in reduced cell size upon growth inhibition. %These results stand in contrast to the model proposed in \cite{bertaux2020}, which predicts that cell size always increases under translation inhibition, which is not consistent with all the available experimental data~\cite{si2017}.

%Discuss the universal growth law suggested by F. Si et al---\\
While the proteome allocation theory is able to capture the deviations in cell size from the nutrient growth law under translation inhibition, its predictions are currently limited to perturbations targeting a few coarse-grained proteome sectors. In recent work, Si \textit{et al.}~\cite{si2017} defined a {\it general growth law} that predicts cell size for a wide range of genetic, antibiotic and nutrient perturbations. Using turbidostat in combination with high-throughput image analysis, Si \textit{et al.} measured the relationship between cell size and growth rate under perturbations to translation, transcription, DNA replication, cell division and cell wall synthesis for a range of nutrient limitations. Their findings 
experimentally confirmed the phenomenological relationship between cell volume $V$, cell cycle duration $\tau_{cyc}$ ($=C+D$ period) and the cell doubling time $\tau$, originally proposed by Donachie~\cite{donachie1968} (Fig.~3E and F):
\begin{equation} \label{eq:si}
V=V_0 2^{\tau_{cyc}/\tau} 
\end{equation}
\noindent where $V_0$ is the cell size at initiation or size of a {\it unit cell}, which remains constant ($V_0\approx 0.27$ $\mu$m$^3$) under growth perturbations. 
Each unit cell contains all the necessary components for self-replication \cite{si2017} such that the cell size at division is the sum of all the unit cells. Interestingly, the volume of a unit cell coincides with a \textit{B. subtilis} bacterial spore volume (0.2 - 0.3 $\mu$m$^3$) --- the smallest self-sufficient bacterial compartment \cite{ojkic2016, lopez2018}. The general growth law for cell size accounts for cell cycles longer than the average doubling time ($\tau_{cyc}>\tau$) where the chromosome contains multiple replication forks (Fig.~3E). After rescaling cell size with the initiation size and growth rate with the inverse of the $C+D$ period, all the available data collapse on the curve predicted by the general growth law (Fig.~3F). The invariance of initiation mass with elongation rate and birth size is consistent with the threshold initiation model, but its mechanistic origin remains unknown. Si \textit{et al.}~\cite{si2017} discuss that the threshold initiation model alone is not sufficient to explain the invariance of unit cell. In addition, the initiator concentration must also be independent of the growth conditions and growth inhibition. Furthermore, they predicted the existence of a specific protein sector that is constant under physiological perturbations that alter the ribosome fraction of the proteome.\\

\noindent {\bf{Coupling between cell shape and growth}}
\begin{figure*}[htp!]
  \includegraphics[width=\linewidth]{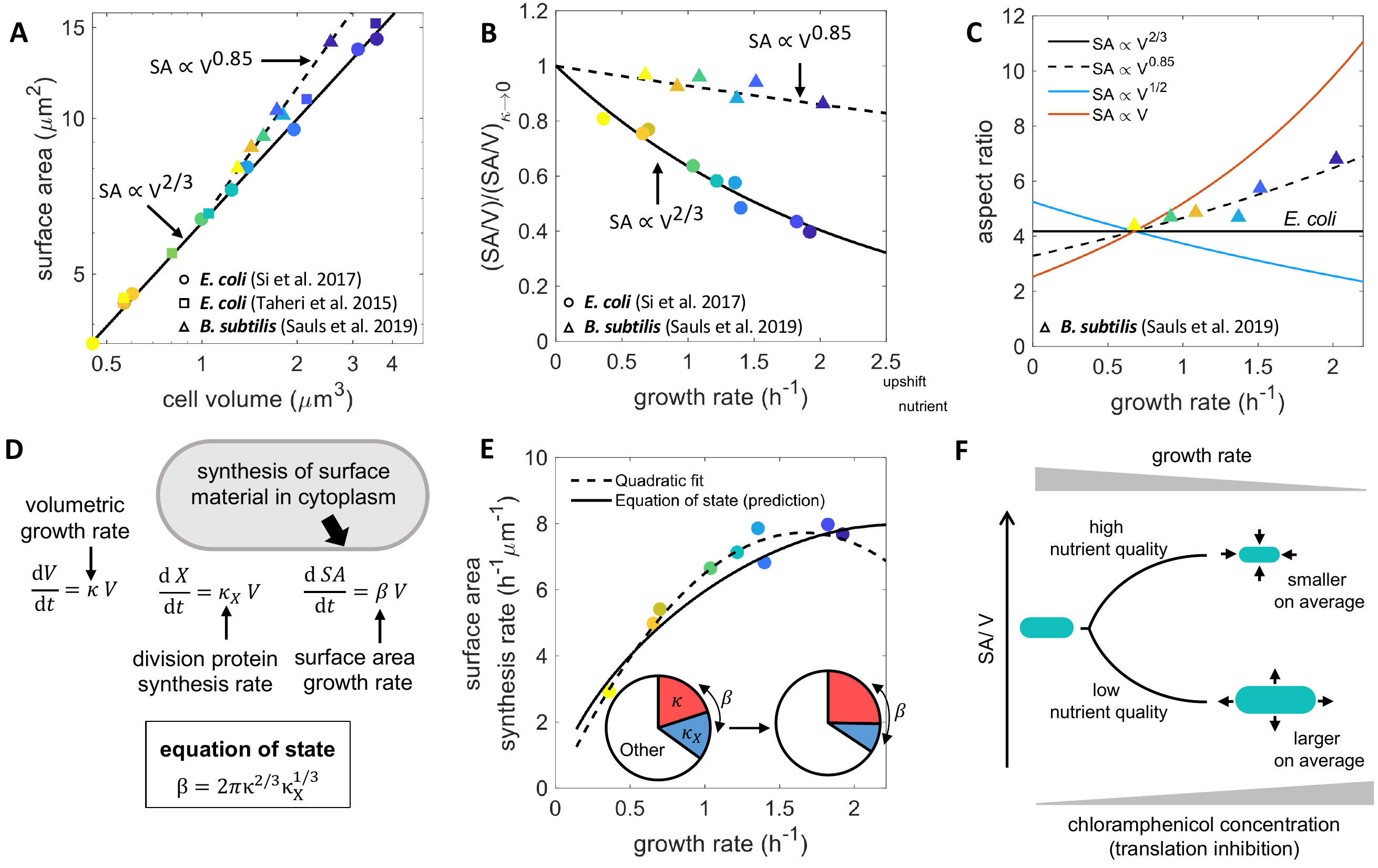}
  \vspace{-10pt}
  \caption{{ \bf Control of cell shape and surface-to-volume ratio.}
{\bf{A.}} Scaling relations between cell surface area and volume for two rod-shaped bacteria, {\it{E. coli}} $SA\propto V^{2/3}$ (solid line) and {\it{B. subtilis}} $SA \propto V^{0.85}$ (dashed line) \cite{ojkic2019,ojkic2021}. 
{\bf{B.}} Surface-to-volume ratio decreases with increasing growth rate: {\it{E. coli}} (solid line) and {\it{B. subtilis}} (dashed line) \cite{ojkic2021}. 
{\bf{C.}} Cell aspect ratio vs growth rate for \textit{B. subtilis} cells grown in different nutrient conditions \cite{sauls2019,ojkic2021}. Using nutrient growth law and various scaling exponents between cell volume and surface area different predictions for aspect ratio are obtained.
{\bf{D.}} Schematic of the Harris and Theriot model~\cite{harris2016} for cell shape control, where cell surface area grows in proportion to cell volume at a rate $\beta$. Harris and Theriot model in conjunction with the threshold initiation model and the surface-to-volume scaling relation ($SA\propto V^{2/3}$) leads to the {\it E. coli} equation of state that connects the rates of cell growth, surface area production and division protein synthesis.
{\bf{E.}} Surface area synthesis rate can be approximated as a quadratic function of growth rate \cite{harris2018} or estimated from the equation of state Eq.~\eqref{eq:beta} \cite{serbanescu2020,ojkic2019}. Inset: resource allocation strategies for the control of cell surface area synthesis.
{\bf{F.}} Bacterial cells treated with translation inhibitory drugs undergo a reduction in active ribosomes, which leads to reduced growth rates in all nutrient conditions. However, the changes in bacterial cell surface-to-volume ratio under translation inhibition are nutrient-quality dependent. In nutrient-poor media surface-to-volume ratio decreases, while cells grown in nutrient-rich media experience an increase in surface-to-volume ratio. 
Data are taken from \cite{si2017,taheri2015,sauls2019}}
  \label{Figure:4}
\end{figure*}

\noindent Together with cell size, the cell shape is an important adaptive trait that is crucial for bacterial growth, motility, nutrient uptake and proliferation \cite{young2006}. Upon nutrient upshift (or downshift), both cell length and width increase (or decrease) with the nutrient-imposed growth rate \cite{zaritsky1973,zaritsky1975,harris2016,si2017,panlilio2021}. However, these geometrical changes occur in a controlled manner in the rod-shaped \textit{E.coli} such that the aspect ratio remains constant~\cite{zaritsky1975,ojkic2019}. In particular, Ojkic \textit{et al.}~\cite{ojkic2019} observed that there is a conserved scaling relation between the surface area ($SA$) and the volume ($V$) of %steady state {\it E. coli} cells
rod-shaped bacterial cells under diverse growth perturbations
\begin{equation} \label{eq:SAVscaling}
SA=\mu V^{\gamma}\;,
\end{equation}
\noindent where $\mu$ is a shape factor related to the aspect ratio of the cell and $\gamma$ is a scaling exponent. In particular, for \textit{E. coli} cells $\gamma=2/3$ and $\mu \approx 2\pi$, suggesting that the aspect ratio of \textit{E. coli} cell is conserved in diverse growth conditions~\cite{ojkic2019}.
%where $\gamma$ is a scaling shape factor related to the aspect ratio of the cell ($\gamma\approx 2\pi$ in {\it E. coli}). 
%of the bacteria prefactor $\gamma$ can be expressed as a function of the cell aspect ratio $\eta=L/W$: $\gamma=\eta \pi \left( \eta \pi/4-\pi/12 \right)^{-2/3}$. 

The scaling between surface area and volume does not obey the $2/3$ exponent for other bacterial species that do not conserve the aspect ratio~\cite{ojkic2021}. For the rod-shaped bacterium \textit{B. subtilis}, a different scaling exponent is observed under growth perturbations $SA\propto V^{0.85}$ (Fig.~4A), implying that the aspect ratio of the cell increases with the growth rate \cite{sharpe1998,sauls2019,ojkic2021}. The scaling between surface area and volume combined with the phenomenological nutrient growth law \cite{Schaechter1958} shows a decrease in surface-to-volume ratio with increasing nutrient-specific growth rate (Fig.~4B) \cite{ojkic2019,ojkic2021}.
%\begin{equation} \label{eq:SAV}
%\frac{SA}{V}\approx 2\pi V_0 ^{-1/3} e^{-\alpha \kappa/3}
%\end{equation}
%\noindent where $V_0$ is the volume at $\kappa=0$ and $\alpha$ is the relative increase in $V$ with $\kappa$. 
Since $SA/V$ decreases with nutrient specific growth rate, this may provide an adaptive mechanism for optimal nutrient uptake in nutrient poor conditions. Furthermore, this relation suggests that bacterial organisms with smaller scaling exponents are highly adaptive across different growth conditions \cite{ojkic2021}.

How is cell surface-to-volume ratio regulated in bacterial cells? Mechanical models of bacterial cell shape control have been proposed~\cite{jiang2010,jiang2011,banerjee2016}, but they do not readily reveal connection between cell shape and growth physiology. Harris and Theriot \cite{harris2016} recently proposed a phenomenological model for $SA/V$ homeostasis, where the relative rates of surface synthesis and volume growth determine cell size (Fig.~4D). In this  model, the surface area of a cell grows in proportion to the cell volume %(Eq.~\ref{eq:V1}):
\begin{equation}
\frac{{\rm d}SA}{{\rm d}t}=\beta V\;,
\end{equation}
where $\beta$ is the volume-specific rate of surface area synthesis. At steady-state, this model leads to the relation $SA/V=\beta/\kappa$. In order to predict cell morphologies under growth rate perturbations, the dependence of the surface area synthesis rate $\beta$ on the growth rate $\kappa$ needs to be specified. To this end, 
%we need to establish how the surface area synthesis rate $\beta$ changes with the nutrient-specific growth rate $\kappa$. 
Harris and Theriot inferred a linear relationship between surface-to-volume ratio and growth rate of \textit{E. coli} cells, using a linear fit to experimental data~\cite{si2017}: $SA/V=-2.8 \kappa+9.3$ \cite{harris2018}. As a result, the surface area synthesis rate $\beta$ is a non-monotonic function of growth rate $\beta=\kappa(-2.8 \kappa+9.3)$ (Fig.~4E). 
%This can be converted to ribosome mass fraction by using the phenomenological relationship $\kappa=\kappa_t(r-r_{min})$ (Fig.~4D):
%\begin{equation} \label{eq:betaHandT}
%\beta=\kappa_t(r-r_{min})(-2.8 \kappa_t(r-r_{min})+9.3)
%\end{equation}

In an alternative approach, combining the surface-to-volume scaling relation (Eq.~\ref{eq:SAVscaling}) with the  threshold initiation model leads to the following {\it equation of state} defining the relationship between the rates $\beta$, $\kappa$ and $\kappa_X$ \cite{ojkic2019,serbanescu2020} (Fig.~4D-E):
\begin{equation} \label{eq:beta}
\beta=\mu \kappa^{\gamma}\kappa_X^{1-\gamma}\;.
%\beta=\eta \pi \left( \frac{\eta \pi}{4}-\frac{\pi}{12} \right)^{-2/3} \kappa_t (r-r_{min}) \left( \frac{\kappa_t(r-r_{min})}{\kappa_p^0 (r_{max}-r)} \right)^{-2/3}
\end{equation}
\noindent In particular, for \textit{E. coli} cells $\beta\approx 2\pi \kappa^{2/3}\kappa_X^{1/3}$ (Fig.~4D). Since $\kappa_X$ decreases with $\kappa$ under nutrient perturbations, the above equation predicts a non-monotonic relation between $\beta$ and $\kappa$ such that $\beta$ increases for $\kappa<\kappa_m$, reaches a maximum at $\kappa=\kappa_m=2\kappa_t(\phi_R^\text{max}-\phi_R^\text{min})/3$, and then decreases for $\kappa>\kappa_m$. The equation of state can be recast in terms of protein mass fractions, $\beta \propto (\phi_R-\phi_R^\text{min})^{2/3}\phi_X^{1/3}$, revealing resource allocation strategies for cell surface area synthesis (Fig.~4E). As hypothesised by Harris and Theriot~\cite{harris2018}, the increasing component of $\beta$ could due to increased availability of raw materials for surface area synthesis, consistent with the proportionality between $\beta$ and $\phi_R$. The decreasing component of $\beta$ could originate from decrease in concentration of cell surface biosynthesis enzymes (belonging to the $X$-sector) with increasing growth rate~\cite{schmidt2016,mori2021}. Under translation inhibition, $\kappa_X$ has a positive correlation with $\kappa$ and thus $\beta$ is an increasing function of $\kappa$. These trends are in quantitative agreement with experimental data~\cite{si2017,serbanescu2020}.%Although the expressions for $\beta$ are not the ideal molecular descriptions, they are extremely helpful in providing information about the cell shape under translation inhibition as they only depend on the ribosome mass fraction (and not on any other geometrical factors such as length or width).

An emerging question in bacterial growth physiology is how quickly bacterial cells respond to growth perturbations and how quickly the cell's morphology adapts to changes in environmental conditions \cite{nguyen2020,panlilio2021}. While the above results for steady-state correlations between cell shape and growth rate are well described by existing theoretical models, it is unclear how the morphological parameters of the cell and the growth rate are coordinated in fast changing conditions. Panlilio \textit{et al.}~\cite{panlilio2021} recently showed that the threshold initiation model of a divisor protein successfully predicts the timescales of cell size adaptation during nutrient upshift in support of the division-centric models of the cell size control even in non-steady growth conditions. In another recent study, Shi \textit{et al.}~\cite{shi2021} investigated how $\beta$ and  $\kappa$ are coupled when stationary cells are switched to fresh medium with a nutrient-limiting capacity. 
The dynamic changes in $SA/V$ along the growth curve are explained by a time delay between $\beta$ and $\kappa$, which originates in the time shift of the expression of cell envelope biosynthesis proteins corresponding to $\beta$ and translational proteins responsible for regulating $\kappa$. Future studies would need to address the molecular mechanisms and pathways for regulating the coupling between cell surface area synthesis and volumetric growth under nutrient modulations, translational and genetic perturbations. Coordination between $\beta$ and $\kappa$ is essential for viable growth and prevention of cell lysis, as observed during drastic cell morphological perturbations induced by antibiotic treatments \cite{nonejuie2013,lamsa2016,htoo2019,ojkic2020,wong2021}.  

While bacterial cells often undergo drastic morphological transformations under growth rate perturbations, %it is unclear
it is an open question how shape transformations contribute to cellular fitness \cite{nonejuie2013,lamsa2016,htoo2019}. Recent findings suggest that cell shape transformation upon translation inhibition promote bacterial growth-rate adaptation in the curved bacterium \textit{C. crescentus} \cite{banerjee2021}. This freshwater organism when exposed to ribosome-targeting antibiotic chloramphenicol becomes wider and more curved \cite{banerjee2021}. Using a mechano-chemical modelling approach, it was shown that the cell envelope softening provides decrease in $SA/V$ to reduce antibiotic influx, while cell curving contributes to increasing cell fitness by promoting faster growth. Cell shape changes could thus provide  adaptive benefits under stress and more work is needed to uncover the feedback between cell morphology and growth in non-steady conditions.  \\

\noindent {\bf{Conclusions and perspectives}}\\
\noindent In this review we present a systems approach to understanding the control of bacterial cell size, shape and growth rate in different nutrient environments. Using the framework of proteome allocation theory developed in recent studies, we identify the relationships between cell morphology and growth parameters and how they are regulated by the proteomic composition of the cell. These studies reveal a key principle of cell size regulation -- a tradeoff between the rates of ribosomal and division protein synthesis sets the optimal cell size in a given nutrient environment. Deviations from this optimal trade-off relationship are observed under translation inhibition. In particular, under translation inhibition in nutrient-rich media, cells invest more ribosomal resources to division than growth, leading to reduced cel size. This is to compensate for a lower translational capacity, which can arise from an increased dilution rate of ribosomes in fast growth conditions, lowering the efficiency of protein synthesis. In nutrient-poor media, cells have a lower nutritional capacity that they compensate by allocating more resources to growth than division protein synthesis, resulting in larger cell sizes. These cell size changes occur with changes in cell surface-to-volume ratio that is reflective of active adaptive response of the cell (Fig.~4F). Decrease in cell volume in nutrient rich media implies a higher surface-to-volume ratio that may increase the influx of nutrients and antibiotics. By contrast, in nutrient poor media, increase in cell volume leads to a lower surface-to-volume ratio that in turn would reduce antibiotic and nutrient influx (Fig.~4F). Therefore, surface-to-volume ratio may play a crucial role in controlling cellular growth by modulating the relative rates of nutrient uptake and antibiotic influx.

%The theoretical models derived from phenomenological relationships between ribosomes, growth rate, cell size and shape provide a useful framework in identifying feedback pathways for the bacterial cell's morphology under nutrient perturbations and translation inhibition. The trade-off between two key protein types: ribosomal proteins and division proteins plays a crucial role in balancing the growth rate and the division protein synthesis rate such that bacterial cells grow to the optimal size given the amount of available nutrients. However ribosome targeting antibiotics inhibit ribosome translation and break the balanced allocation of resources. Bacterial cells successfully maintain their shapes, but the cell size changes in a nutrient-dependent manner suggesting an interesting adaptation mechanism 
%The latter .

Coarse-grained proteome partitioning models \cite{scott2010,scott2014} come with the challenge of identifying the proteome sector to which the protein of interest belongs or if a new sector has to be included. In recent work, it has been suggested that the division proteins belong to the non-ribosomal $P$-sector~\cite{taheri2015,si2017,panlilio2021}, while others considered a separate proteome sector for division proteins \cite{basan2015,bertaux2020,serbanescu2020}. By accounting a separate division protein sector, Serbanescu \textit{et al.} \cite{serbanescu2020} showed that the division protein sector behaves as a sub-class of $P$-sector proteins, given there is a negative correlation between protein mass fractions of $X$-sector and $R$-sector proteins under nutrient perturbations. Existing data support this model and also exclude other possible models of division proteins belonging to the $R$ or $Q$ sectors. For instance, if the division molecule is in the $R$-sector, then we expect $\kappa_X$ to be positively correlated with ribosome mass fraction at all conditions. Therefore, under translation inhibition we expect $\kappa_X$ to increase. Since growth rate decreases under translation inhibition, we would predict cell volume to decrease for all conditions. This is inconsistent with experimental data \cite{si2017}. However, if the division proteins are in the $Q$-sector, then the invariance of $Q$-sector with nutrient shifts and translation inhibition would imply a constant division protein synthesis rate for all conditions. Therefore cell size would always decrease under translation inhibition, inconsistent with experimental data. Hence the division protein sector needs to be in the non-ribosomal $P$-sector to capture the tradeoff between the rates of growth and division protein synthesis under nutrient perturbations.

The phenomenological growth laws that allow us to predict how bacterial cells regulate their grow rates, proteome composition, cell shape and size are well established for steady-state growth conditions. We currently have a limited understanding of the dynamics of cellular state variables in non-steady state conditions. With recent developments in high-throughput single-cell studies of bacterial growth, understanding cell behaviors during transition from one growth condition to another is of great interest~\cite{mori2017,erickson2017,basan2020,bakshi2021,shi2021,harris2016,banerjee2021}. Recent single-cell studies show that during single nutrient shifts \cite{harris2016,panlilio2021} or antibiotic treatments~\cite{banerjee2021} bacterial cells preserve their size control strategies, indicating that the division proteins are actively regulated in changing environments. When subject to nutrient fluctuations between low and high concentrations of nutrients \cite{nguyen2020}, cells adopt a growth rate that cannot be predicted from the individual growth rates in poor and rich nutrient conditions
%the low- and high- nutrient conditions 
as predicted by the Monod's law (Eq.~\ref{eq:monod}), indicating that the relationship between ribosomes and growth rate does not hold in non-steady conditions. This adaptive mechanism leads to an interesting hypothesis that has been previously formulated in the context of fast transition to a new environmental condition. For instance, bacterial cells have a ribosome reserve arising from the inactive ribosomes that do not participate in translation. This inactive fraction allows cells to rapidly produce proteins in optimal amounts for the new growth condition \cite{mori2017,ni2020}. How resource allocation strategies are achieved during fast changing environments and whether growth-division tradeoff model still holds for non-steady fluctuating environments remains an open question. \\

%{\color{RoyalBlue} Another interesting question still to be answered is how combinations of translation inhibitory drugs affect the size and shape of bacteria. When simultaneously applying two antibiotics, the resulting effect can be compared to the effect of the individual antibiotics resulting in {\it{antagonistic}} (combined effect is weaker) or {\it{synergistic}} (combined effect is stronger) interactions.  Kav{\v{c}}i{\v{c}}, Tka{\v{c}}ik and Bollenbach \cite{kavvcivc2021} have investigated the interaction network between widely used antibiotics inhibiting ribosome translation. Intuitively, a couple of antibiotics with similar mode of action should have an additive effect, decreasing the growth rate of bacterial cells more than the individual drugs alone. However, suppresive drug interactions are a result of ``ribosome traffic jams'' and translation bottlenecks. Although the response of growth rate to the cocktail of antibiotics can be explained with biophysical models \cite{kavvcivc2020} which gives the synergistic or antagonistic behavior, the cell morphology is not simply characterized and requires both cell shape information and cycle progression component (division proteins).\\

\noindent {\bf References}
\bibliography{References}{}

%merlin.mbs apsrev4-1.bst 2010-07-25 4.21a (PWD, AO, DPC) hacked
%Control: key (0)
%Control: author (8) initials jnrlst
%Control: editor formatted (1) identically to author
%Control: production of article title (-1) disabled
%Control: page (0) single
%Control: year (1) truncated
%Control: production of eprint (0) enabled
\begin{thebibliography}{98}%
\makeatletter
\providecommand \@ifxundefined [1]{%
 \@ifx{#1\undefined}
}%
\providecommand \@ifnum [1]{%
 \ifnum #1\expandafter \@firstoftwo
 \else \expandafter \@secondoftwo
 \fi
}%
\providecommand \@ifx [1]{%
 \ifx #1\expandafter \@firstoftwo
 \else \expandafter \@secondoftwo
 \fi
}%
\providecommand \natexlab [1]{#1}%
\providecommand \enquote  [1]{``#1''}%
\providecommand \bibnamefont  [1]{#1}%
\providecommand \bibfnamefont [1]{#1}%
\providecommand \citenamefont [1]{#1}%
\providecommand \href@noop [0]{\@secondoftwo}%
\providecommand \href [0]{\begingroup \@sanitize@url \@href}%
\providecommand \@href[1]{\@@startlink{#1}\@@href}%
\providecommand \@@href[1]{\endgroup#1\@@endlink}%
\providecommand \@sanitize@url [0]{\catcode `\\12\catcode `\$12\catcode
  `\&12\catcode `\#12\catcode `\^12\catcode `\_12\catcode `\%12\relax}%
\providecommand \@@startlink[1]{}%
\providecommand \@@endlink[0]{}%
\providecommand \url  [0]{\begingroup\@sanitize@url \@url }%
\providecommand \@url [1]{\endgroup\@href {#1}{\urlprefix }}%
\providecommand \urlprefix  [0]{URL }%
\providecommand \Eprint [0]{\href }%
\providecommand \doibase [0]{http://dx.doi.org/}%
\providecommand \selectlanguage [0]{\@gobble}%
\providecommand \bibinfo  [0]{\@secondoftwo}%
\providecommand \bibfield  [0]{\@secondoftwo}%
\providecommand \translation [1]{[#1]}%
\providecommand \BibitemOpen [0]{}%
\providecommand \bibitemStop [0]{}%
\providecommand \bibitemNoStop [0]{.\EOS\space}%
\providecommand \EOS [0]{\spacefactor3000\relax}%
\providecommand \BibitemShut  [1]{\csname bibitem#1\endcsname}%
\let\auto@bib@innerbib\@empty
%</preamble>
\bibitem [{\citenamefont {Young}(2006)}]{young2006}%
  \BibitemOpen
  \bibfield  {author} {\bibinfo {author} {\bibfnamefont {K.~D.}\ \bibnamefont
  {Young}},\ }\href@noop {} {\bibfield  {journal} {\bibinfo  {journal}
  {Microbiology and molecular biology reviews}\ }\textbf {\bibinfo {volume}
  {70}},\ \bibinfo {pages} {660} (\bibinfo {year} {2006})}\BibitemShut
  {NoStop}%
\bibitem [{\citenamefont {Schaechter}\ \emph {et~al.}(1958)\citenamefont
  {Schaechter}, \citenamefont {Maal{\o}e},\ and\ \citenamefont
  {Kjeldgaard}}]{Schaechter1958}%
  \BibitemOpen
  \bibfield  {author} {\bibinfo {author} {\bibfnamefont {M.}~\bibnamefont
  {Schaechter}}, \bibinfo {author} {\bibfnamefont {O.}~\bibnamefont
  {Maal{\o}e}}, \ and\ \bibinfo {author} {\bibfnamefont {N.~O.}\ \bibnamefont
  {Kjeldgaard}},\ }\href@noop {} {\bibfield  {journal} {\bibinfo  {journal}
  {Microbiology}\ }\textbf {\bibinfo {volume} {19}},\ \bibinfo {pages} {592}
  (\bibinfo {year} {1958})}\BibitemShut {NoStop}%
\bibitem [{\citenamefont {Taheri-Araghi}\ \emph
  {et~al.}(2015{\natexlab{a}})\citenamefont {Taheri-Araghi}, \citenamefont
  {Bradde}, \citenamefont {Sauls}, \citenamefont {Hill}, \citenamefont {Levin},
  \citenamefont {Paulsson}, \citenamefont {Vergassola},\ and\ \citenamefont
  {Jun}}]{taheri2015}%
  \BibitemOpen
  \bibfield  {author} {\bibinfo {author} {\bibfnamefont {S.}~\bibnamefont
  {Taheri-Araghi}}, \bibinfo {author} {\bibfnamefont {S.}~\bibnamefont
  {Bradde}}, \bibinfo {author} {\bibfnamefont {J.~T.}\ \bibnamefont {Sauls}},
  \bibinfo {author} {\bibfnamefont {N.~S.}\ \bibnamefont {Hill}}, \bibinfo
  {author} {\bibfnamefont {P.~A.}\ \bibnamefont {Levin}}, \bibinfo {author}
  {\bibfnamefont {J.}~\bibnamefont {Paulsson}}, \bibinfo {author}
  {\bibfnamefont {M.}~\bibnamefont {Vergassola}}, \ and\ \bibinfo {author}
  {\bibfnamefont {S.}~\bibnamefont {Jun}},\ }\href@noop {} {\bibfield
  {journal} {\bibinfo  {journal} {Current biology}\ }\textbf {\bibinfo {volume}
  {25}},\ \bibinfo {pages} {385} (\bibinfo {year}
  {2015}{\natexlab{a}})}\BibitemShut {NoStop}%
\bibitem [{\citenamefont {Vadia}\ and\ \citenamefont
  {Levin}(2015)}]{vadia2015}%
  \BibitemOpen
  \bibfield  {author} {\bibinfo {author} {\bibfnamefont {S.}~\bibnamefont
  {Vadia}}\ and\ \bibinfo {author} {\bibfnamefont {P.~A.}\ \bibnamefont
  {Levin}},\ }\href@noop {} {\bibfield  {journal} {\bibinfo  {journal} {Current
  opinion in microbiology}\ }\textbf {\bibinfo {volume} {24}},\ \bibinfo
  {pages} {96} (\bibinfo {year} {2015})}\BibitemShut {NoStop}%
\bibitem [{\citenamefont {Si}\ \emph {et~al.}(2017)\citenamefont {Si},
  \citenamefont {Li}, \citenamefont {Cox}, \citenamefont {Sauls}, \citenamefont
  {Azizi}, \citenamefont {Sou}, \citenamefont {Schwartz}, \citenamefont
  {Erickstad}, \citenamefont {Jun}, \citenamefont {Li} \emph
  {et~al.}}]{si2017}%
  \BibitemOpen
  \bibfield  {author} {\bibinfo {author} {\bibfnamefont {F.}~\bibnamefont
  {Si}}, \bibinfo {author} {\bibfnamefont {D.}~\bibnamefont {Li}}, \bibinfo
  {author} {\bibfnamefont {S.~E.}\ \bibnamefont {Cox}}, \bibinfo {author}
  {\bibfnamefont {J.~T.}\ \bibnamefont {Sauls}}, \bibinfo {author}
  {\bibfnamefont {O.}~\bibnamefont {Azizi}}, \bibinfo {author} {\bibfnamefont
  {C.}~\bibnamefont {Sou}}, \bibinfo {author} {\bibfnamefont {A.~B.}\
  \bibnamefont {Schwartz}}, \bibinfo {author} {\bibfnamefont {M.~J.}\
  \bibnamefont {Erickstad}}, \bibinfo {author} {\bibfnamefont {Y.}~\bibnamefont
  {Jun}}, \bibinfo {author} {\bibfnamefont {X.}~\bibnamefont {Li}},  \emph
  {et~al.},\ }\href@noop {} {\bibfield  {journal} {\bibinfo  {journal} {Current
  Biology}\ }\textbf {\bibinfo {volume} {27}},\ \bibinfo {pages} {1278}
  (\bibinfo {year} {2017})}\BibitemShut {NoStop}%
\bibitem [{\citenamefont {Dai}\ \emph {et~al.}(2018)\citenamefont {Dai},
  \citenamefont {Shen}, \citenamefont {Wang},\ and\ \citenamefont
  {Zhu}}]{dai2018}%
  \BibitemOpen
  \bibfield  {author} {\bibinfo {author} {\bibfnamefont {X.}~\bibnamefont
  {Dai}}, \bibinfo {author} {\bibfnamefont {Z.}~\bibnamefont {Shen}}, \bibinfo
  {author} {\bibfnamefont {Y.}~\bibnamefont {Wang}}, \ and\ \bibinfo {author}
  {\bibfnamefont {M.}~\bibnamefont {Zhu}},\ }\href@noop {} {\bibfield
  {journal} {\bibinfo  {journal} {MSphere}\ }\textbf {\bibinfo {volume} {3}}
  (\bibinfo {year} {2018})}\BibitemShut {NoStop}%
\bibitem [{\citenamefont {Sauls}\ \emph {et~al.}(2019)\citenamefont {Sauls},
  \citenamefont {Cox}, \citenamefont {Do}, \citenamefont {Castillo},
  \citenamefont {Ghulam-Jelani},\ and\ \citenamefont {Jun}}]{sauls2019}%
  \BibitemOpen
  \bibfield  {author} {\bibinfo {author} {\bibfnamefont {J.~T.}\ \bibnamefont
  {Sauls}}, \bibinfo {author} {\bibfnamefont {S.~E.}\ \bibnamefont {Cox}},
  \bibinfo {author} {\bibfnamefont {Q.}~\bibnamefont {Do}}, \bibinfo {author}
  {\bibfnamefont {V.}~\bibnamefont {Castillo}}, \bibinfo {author}
  {\bibfnamefont {Z.}~\bibnamefont {Ghulam-Jelani}}, \ and\ \bibinfo {author}
  {\bibfnamefont {S.}~\bibnamefont {Jun}},\ }\href@noop {} {\bibfield
  {journal} {\bibinfo  {journal} {Mbio}\ }\textbf {\bibinfo {volume} {10}}
  (\bibinfo {year} {2019})}\BibitemShut {NoStop}%
\bibitem [{\citenamefont {Basan}\ \emph {et~al.}(2015)\citenamefont {Basan},
  \citenamefont {Zhu}, \citenamefont {Dai}, \citenamefont {Warren},
  \citenamefont {S{\'e}vin}, \citenamefont {Wang},\ and\ \citenamefont
  {Hwa}}]{basan2015}%
  \BibitemOpen
  \bibfield  {author} {\bibinfo {author} {\bibfnamefont {M.}~\bibnamefont
  {Basan}}, \bibinfo {author} {\bibfnamefont {M.}~\bibnamefont {Zhu}}, \bibinfo
  {author} {\bibfnamefont {X.}~\bibnamefont {Dai}}, \bibinfo {author}
  {\bibfnamefont {M.}~\bibnamefont {Warren}}, \bibinfo {author} {\bibfnamefont
  {D.}~\bibnamefont {S{\'e}vin}}, \bibinfo {author} {\bibfnamefont {Y.-P.}\
  \bibnamefont {Wang}}, \ and\ \bibinfo {author} {\bibfnamefont
  {T.}~\bibnamefont {Hwa}},\ }\href@noop {} {\bibfield  {journal} {\bibinfo
  {journal} {Molecular systems biology}\ }\textbf {\bibinfo {volume} {11}},\
  \bibinfo {pages} {836} (\bibinfo {year} {2015})}\BibitemShut {NoStop}%
\bibitem [{\citenamefont {Amir}(2014)}]{amir2014}%
  \BibitemOpen
  \bibfield  {author} {\bibinfo {author} {\bibfnamefont {A.}~\bibnamefont
  {Amir}},\ }\href@noop {} {\bibfield  {journal} {\bibinfo  {journal} {Physical
  review letters}\ }\textbf {\bibinfo {volume} {112}},\ \bibinfo {pages}
  {208102} (\bibinfo {year} {2014})}\BibitemShut {NoStop}%
\bibitem [{\citenamefont {Banerjee}\ \emph {et~al.}(2017)\citenamefont
  {Banerjee}, \citenamefont {Lo}, \citenamefont {Daddysman}, \citenamefont
  {Selewa}, \citenamefont {Kuntz}, \citenamefont {Dinner},\ and\ \citenamefont
  {Scherer}}]{banerjee2017}%
  \BibitemOpen
  \bibfield  {author} {\bibinfo {author} {\bibfnamefont {S.}~\bibnamefont
  {Banerjee}}, \bibinfo {author} {\bibfnamefont {K.}~\bibnamefont {Lo}},
  \bibinfo {author} {\bibfnamefont {M.~K.}\ \bibnamefont {Daddysman}}, \bibinfo
  {author} {\bibfnamefont {A.}~\bibnamefont {Selewa}}, \bibinfo {author}
  {\bibfnamefont {T.}~\bibnamefont {Kuntz}}, \bibinfo {author} {\bibfnamefont
  {A.~R.}\ \bibnamefont {Dinner}}, \ and\ \bibinfo {author} {\bibfnamefont
  {N.~F.}\ \bibnamefont {Scherer}},\ }\href@noop {} {\bibfield  {journal}
  {\bibinfo  {journal} {Nature microbiology}\ }\textbf {\bibinfo {volume}
  {2}},\ \bibinfo {pages} {1} (\bibinfo {year} {2017})}\BibitemShut {NoStop}%
\bibitem [{\citenamefont {Wallden}\ \emph {et~al.}(2016)\citenamefont
  {Wallden}, \citenamefont {Fange}, \citenamefont {Lundius}, \citenamefont
  {Baltekin},\ and\ \citenamefont {Elf}}]{wallden2016}%
  \BibitemOpen
  \bibfield  {author} {\bibinfo {author} {\bibfnamefont {M.}~\bibnamefont
  {Wallden}}, \bibinfo {author} {\bibfnamefont {D.}~\bibnamefont {Fange}},
  \bibinfo {author} {\bibfnamefont {E.~G.}\ \bibnamefont {Lundius}}, \bibinfo
  {author} {\bibfnamefont {{\"O}.}~\bibnamefont {Baltekin}}, \ and\ \bibinfo
  {author} {\bibfnamefont {J.}~\bibnamefont {Elf}},\ }\href@noop {} {\bibfield
  {journal} {\bibinfo  {journal} {Cell}\ }\textbf {\bibinfo {volume} {166}},\
  \bibinfo {pages} {729} (\bibinfo {year} {2016})}\BibitemShut {NoStop}%
\bibitem [{\citenamefont {Cadart}\ \emph {et~al.}(2019)\citenamefont {Cadart},
  \citenamefont {Venkova}, \citenamefont {Recho}, \citenamefont {Lagomarsino},\
  and\ \citenamefont {Piel}}]{cadart2019}%
  \BibitemOpen
  \bibfield  {author} {\bibinfo {author} {\bibfnamefont {C.}~\bibnamefont
  {Cadart}}, \bibinfo {author} {\bibfnamefont {L.}~\bibnamefont {Venkova}},
  \bibinfo {author} {\bibfnamefont {P.}~\bibnamefont {Recho}}, \bibinfo
  {author} {\bibfnamefont {M.~C.}\ \bibnamefont {Lagomarsino}}, \ and\ \bibinfo
  {author} {\bibfnamefont {M.}~\bibnamefont {Piel}},\ }\href@noop {} {\bibfield
   {journal} {\bibinfo  {journal} {Nature Physics}\ }\textbf {\bibinfo {volume}
  {15}},\ \bibinfo {pages} {993} (\bibinfo {year} {2019})}\BibitemShut
  {NoStop}%
\bibitem [{\citenamefont {Campos}\ \emph {et~al.}(2014)\citenamefont {Campos},
  \citenamefont {Surovtsev}, \citenamefont {Kato}, \citenamefont {Paintdakhi},
  \citenamefont {Beltran}, \citenamefont {Ebmeier},\ and\ \citenamefont
  {Jacobs-Wagner}}]{campos2014}%
  \BibitemOpen
  \bibfield  {author} {\bibinfo {author} {\bibfnamefont {M.}~\bibnamefont
  {Campos}}, \bibinfo {author} {\bibfnamefont {I.~V.}\ \bibnamefont
  {Surovtsev}}, \bibinfo {author} {\bibfnamefont {S.}~\bibnamefont {Kato}},
  \bibinfo {author} {\bibfnamefont {A.}~\bibnamefont {Paintdakhi}}, \bibinfo
  {author} {\bibfnamefont {B.}~\bibnamefont {Beltran}}, \bibinfo {author}
  {\bibfnamefont {S.~E.}\ \bibnamefont {Ebmeier}}, \ and\ \bibinfo {author}
  {\bibfnamefont {C.}~\bibnamefont {Jacobs-Wagner}},\ }\href@noop {} {\bibfield
   {journal} {\bibinfo  {journal} {Cell}\ }\textbf {\bibinfo {volume} {159}},\
  \bibinfo {pages} {1433} (\bibinfo {year} {2014})}\BibitemShut {NoStop}%
\bibitem [{\citenamefont {Deforet}\ \emph {et~al.}(2015)\citenamefont
  {Deforet}, \citenamefont {Van~Ditmarsch},\ and\ \citenamefont
  {Xavier}}]{deforet2015}%
  \BibitemOpen
  \bibfield  {author} {\bibinfo {author} {\bibfnamefont {M.}~\bibnamefont
  {Deforet}}, \bibinfo {author} {\bibfnamefont {D.}~\bibnamefont
  {Van~Ditmarsch}}, \ and\ \bibinfo {author} {\bibfnamefont {J.~B.}\
  \bibnamefont {Xavier}},\ }\href@noop {} {\bibfield  {journal} {\bibinfo
  {journal} {Biophysical journal}\ }\textbf {\bibinfo {volume} {109}},\
  \bibinfo {pages} {521} (\bibinfo {year} {2015})}\BibitemShut {NoStop}%
\bibitem [{\citenamefont {Jun}\ and\ \citenamefont
  {Taheri-Araghi}(2015)}]{jun2015}%
  \BibitemOpen
  \bibfield  {author} {\bibinfo {author} {\bibfnamefont {S.}~\bibnamefont
  {Jun}}\ and\ \bibinfo {author} {\bibfnamefont {S.}~\bibnamefont
  {Taheri-Araghi}},\ }\href@noop {} {\bibfield  {journal} {\bibinfo  {journal}
  {Trends in microbiology}\ }\textbf {\bibinfo {volume} {23}},\ \bibinfo
  {pages} {4} (\bibinfo {year} {2015})}\BibitemShut {NoStop}%
\bibitem [{\citenamefont {Jun}\ \emph {et~al.}(2018)\citenamefont {Jun},
  \citenamefont {Si}, \citenamefont {Pugatch},\ and\ \citenamefont
  {Scott}}]{jun2018}%
  \BibitemOpen
  \bibfield  {author} {\bibinfo {author} {\bibfnamefont {S.}~\bibnamefont
  {Jun}}, \bibinfo {author} {\bibfnamefont {F.}~\bibnamefont {Si}}, \bibinfo
  {author} {\bibfnamefont {R.}~\bibnamefont {Pugatch}}, \ and\ \bibinfo
  {author} {\bibfnamefont {M.}~\bibnamefont {Scott}},\ }\href@noop {}
  {\bibfield  {journal} {\bibinfo  {journal} {Reports on Progress in Physics}\
  }\textbf {\bibinfo {volume} {81}},\ \bibinfo {pages} {056601} (\bibinfo
  {year} {2018})}\BibitemShut {NoStop}%
\bibitem [{\citenamefont {Osella}\ \emph {et~al.}(2014)\citenamefont {Osella},
  \citenamefont {Nugent},\ and\ \citenamefont {Lagomarsino}}]{osella2014}%
  \BibitemOpen
  \bibfield  {author} {\bibinfo {author} {\bibfnamefont {M.}~\bibnamefont
  {Osella}}, \bibinfo {author} {\bibfnamefont {E.}~\bibnamefont {Nugent}}, \
  and\ \bibinfo {author} {\bibfnamefont {M.~C.}\ \bibnamefont {Lagomarsino}},\
  }\href@noop {} {\bibfield  {journal} {\bibinfo  {journal} {Proceedings of the
  National Academy of Sciences}\ }\textbf {\bibinfo {volume} {111}},\ \bibinfo
  {pages} {3431} (\bibinfo {year} {2014})}\BibitemShut {NoStop}%
\bibitem [{\citenamefont {Donachie}(1968)}]{donachie1968}%
  \BibitemOpen
  \bibfield  {author} {\bibinfo {author} {\bibfnamefont {W.~D.}\ \bibnamefont
  {Donachie}},\ }\href@noop {} {\bibfield  {journal} {\bibinfo  {journal}
  {Nature}\ }\textbf {\bibinfo {volume} {219}},\ \bibinfo {pages} {1077}
  (\bibinfo {year} {1968})}\BibitemShut {NoStop}%
\bibitem [{\citenamefont {Ho}\ and\ \citenamefont {Amir}(2015)}]{ho2015}%
  \BibitemOpen
  \bibfield  {author} {\bibinfo {author} {\bibfnamefont {P.-Y.}\ \bibnamefont
  {Ho}}\ and\ \bibinfo {author} {\bibfnamefont {A.}~\bibnamefont {Amir}},\
  }\href@noop {} {\bibfield  {journal} {\bibinfo  {journal} {Frontiers in
  microbiology}\ }\textbf {\bibinfo {volume} {6}},\ \bibinfo {pages} {662}
  (\bibinfo {year} {2015})}\BibitemShut {NoStop}%
\bibitem [{\citenamefont {Ghusinga}\ \emph {et~al.}(2016)\citenamefont
  {Ghusinga}, \citenamefont {Vargas-Garcia},\ and\ \citenamefont
  {Singh}}]{ghusinga2016}%
  \BibitemOpen
  \bibfield  {author} {\bibinfo {author} {\bibfnamefont {K.~R.}\ \bibnamefont
  {Ghusinga}}, \bibinfo {author} {\bibfnamefont {C.~A.}\ \bibnamefont
  {Vargas-Garcia}}, \ and\ \bibinfo {author} {\bibfnamefont {A.}~\bibnamefont
  {Singh}},\ }\href@noop {} {\bibfield  {journal} {\bibinfo  {journal}
  {Scientific reports}\ }\textbf {\bibinfo {volume} {6}},\ \bibinfo {pages} {1}
  (\bibinfo {year} {2016})}\BibitemShut {NoStop}%
\bibitem [{\citenamefont {Harris}\ and\ \citenamefont
  {Theriot}(2016)}]{harris2016}%
  \BibitemOpen
  \bibfield  {author} {\bibinfo {author} {\bibfnamefont {L.~K.}\ \bibnamefont
  {Harris}}\ and\ \bibinfo {author} {\bibfnamefont {J.~A.}\ \bibnamefont
  {Theriot}},\ }\href@noop {} {\bibfield  {journal} {\bibinfo  {journal}
  {Cell}\ }\textbf {\bibinfo {volume} {165}},\ \bibinfo {pages} {1479}
  (\bibinfo {year} {2016})}\BibitemShut {NoStop}%
\bibitem [{\citenamefont {Si}\ \emph {et~al.}(2019)\citenamefont {Si},
  \citenamefont {Le~Treut}, \citenamefont {Sauls}, \citenamefont {Vadia},
  \citenamefont {Levin},\ and\ \citenamefont {Jun}}]{si2019}%
  \BibitemOpen
  \bibfield  {author} {\bibinfo {author} {\bibfnamefont {F.}~\bibnamefont
  {Si}}, \bibinfo {author} {\bibfnamefont {G.}~\bibnamefont {Le~Treut}},
  \bibinfo {author} {\bibfnamefont {J.~T.}\ \bibnamefont {Sauls}}, \bibinfo
  {author} {\bibfnamefont {S.}~\bibnamefont {Vadia}}, \bibinfo {author}
  {\bibfnamefont {P.~A.}\ \bibnamefont {Levin}}, \ and\ \bibinfo {author}
  {\bibfnamefont {S.}~\bibnamefont {Jun}},\ }\href@noop {} {\bibfield
  {journal} {\bibinfo  {journal} {Current Biology}\ }\textbf {\bibinfo {volume}
  {29}},\ \bibinfo {pages} {1760} (\bibinfo {year} {2019})}\BibitemShut
  {NoStop}%
\bibitem [{\citenamefont {Serbanescu}\ \emph {et~al.}(2020)\citenamefont
  {Serbanescu}, \citenamefont {Ojkic},\ and\ \citenamefont
  {Banerjee}}]{serbanescu2020}%
  \BibitemOpen
  \bibfield  {author} {\bibinfo {author} {\bibfnamefont {D.}~\bibnamefont
  {Serbanescu}}, \bibinfo {author} {\bibfnamefont {N.}~\bibnamefont {Ojkic}}, \
  and\ \bibinfo {author} {\bibfnamefont {S.}~\bibnamefont {Banerjee}},\
  }\href@noop {} {\bibfield  {journal} {\bibinfo  {journal} {Cell Reports}\
  }\textbf {\bibinfo {volume} {32}},\ \bibinfo {pages} {108183} (\bibinfo
  {year} {2020})}\BibitemShut {NoStop}%
\bibitem [{\citenamefont {Panlilio}\ \emph {et~al.}(2021)\citenamefont
  {Panlilio}, \citenamefont {Grilli}, \citenamefont {Tallarico}, \citenamefont
  {Iuliani}, \citenamefont {Sclavi}, \citenamefont {Cicuta},\ and\
  \citenamefont {Lagomarsino}}]{panlilio2021}%
  \BibitemOpen
  \bibfield  {author} {\bibinfo {author} {\bibfnamefont {M.}~\bibnamefont
  {Panlilio}}, \bibinfo {author} {\bibfnamefont {J.}~\bibnamefont {Grilli}},
  \bibinfo {author} {\bibfnamefont {G.}~\bibnamefont {Tallarico}}, \bibinfo
  {author} {\bibfnamefont {I.}~\bibnamefont {Iuliani}}, \bibinfo {author}
  {\bibfnamefont {B.}~\bibnamefont {Sclavi}}, \bibinfo {author} {\bibfnamefont
  {P.}~\bibnamefont {Cicuta}}, \ and\ \bibinfo {author} {\bibfnamefont {M.~C.}\
  \bibnamefont {Lagomarsino}},\ }\href@noop {} {\bibfield  {journal} {\bibinfo
  {journal} {Proceedings of the National Academy of Sciences}\ }\textbf
  {\bibinfo {volume} {118}} (\bibinfo {year} {2021})}\BibitemShut {NoStop}%
\bibitem [{\citenamefont {Ojkic}\ \emph {et~al.}(2019)\citenamefont {Ojkic},
  \citenamefont {Serbanescu},\ and\ \citenamefont {Banerjee}}]{ojkic2019}%
  \BibitemOpen
  \bibfield  {author} {\bibinfo {author} {\bibfnamefont {N.}~\bibnamefont
  {Ojkic}}, \bibinfo {author} {\bibfnamefont {D.}~\bibnamefont {Serbanescu}}, \
  and\ \bibinfo {author} {\bibfnamefont {S.}~\bibnamefont {Banerjee}},\
  }\href@noop {} {\bibfield  {journal} {\bibinfo  {journal} {Elife}\ }\textbf
  {\bibinfo {volume} {8}},\ \bibinfo {pages} {e47033} (\bibinfo {year}
  {2019})}\BibitemShut {NoStop}%
\bibitem [{\citenamefont {Scott}\ and\ \citenamefont {Hwa}(2011)}]{scott2011}%
  \BibitemOpen
  \bibfield  {author} {\bibinfo {author} {\bibfnamefont {M.}~\bibnamefont
  {Scott}}\ and\ \bibinfo {author} {\bibfnamefont {T.}~\bibnamefont {Hwa}},\
  }\href@noop {} {\bibfield  {journal} {\bibinfo  {journal} {Current Opinion in
  Biotechnology}\ }\textbf {\bibinfo {volume} {22}},\ \bibinfo {pages} {559}
  (\bibinfo {year} {2011})}\BibitemShut {NoStop}%
\bibitem [{\citenamefont {Scott}\ \emph {et~al.}(2014)\citenamefont {Scott},
  \citenamefont {Klumpp}, \citenamefont {Mateescu},\ and\ \citenamefont
  {Hwa}}]{scott2014}%
  \BibitemOpen
  \bibfield  {author} {\bibinfo {author} {\bibfnamefont {M.}~\bibnamefont
  {Scott}}, \bibinfo {author} {\bibfnamefont {S.}~\bibnamefont {Klumpp}},
  \bibinfo {author} {\bibfnamefont {E.~M.}\ \bibnamefont {Mateescu}}, \ and\
  \bibinfo {author} {\bibfnamefont {T.}~\bibnamefont {Hwa}},\ }\href@noop {}
  {\bibfield  {journal} {\bibinfo  {journal} {Molecular systems biology}\
  }\textbf {\bibinfo {volume} {10}},\ \bibinfo {pages} {747} (\bibinfo {year}
  {2014})}\BibitemShut {NoStop}%
\bibitem [{\citenamefont {Klumpp}\ and\ \citenamefont
  {Hwa}(2014)}]{klumpp2014}%
  \BibitemOpen
  \bibfield  {author} {\bibinfo {author} {\bibfnamefont {S.}~\bibnamefont
  {Klumpp}}\ and\ \bibinfo {author} {\bibfnamefont {T.}~\bibnamefont {Hwa}},\
  }\href@noop {} {\bibfield  {journal} {\bibinfo  {journal} {Current Opinion in
  Biotechnology}\ }\textbf {\bibinfo {volume} {28}},\ \bibinfo {pages} {96}
  (\bibinfo {year} {2014})}\BibitemShut {NoStop}%
\bibitem [{\citenamefont {Taheri-Araghi}\ \emph
  {et~al.}(2015{\natexlab{b}})\citenamefont {Taheri-Araghi}, \citenamefont
  {Brown}, \citenamefont {Sauls}, \citenamefont {McIntosh},\ and\ \citenamefont
  {Jun}}]{taheri2015b}%
  \BibitemOpen
  \bibfield  {author} {\bibinfo {author} {\bibfnamefont {S.}~\bibnamefont
  {Taheri-Araghi}}, \bibinfo {author} {\bibfnamefont {S.~D.}\ \bibnamefont
  {Brown}}, \bibinfo {author} {\bibfnamefont {J.~T.}\ \bibnamefont {Sauls}},
  \bibinfo {author} {\bibfnamefont {D.~B.}\ \bibnamefont {McIntosh}}, \ and\
  \bibinfo {author} {\bibfnamefont {S.}~\bibnamefont {Jun}},\ }\href@noop {}
  {\bibfield  {journal} {\bibinfo  {journal} {Annual Review of Biophysics}\
  }\textbf {\bibinfo {volume} {44}},\ \bibinfo {pages} {123} (\bibinfo {year}
  {2015}{\natexlab{b}})}\BibitemShut {NoStop}%
\bibitem [{\citenamefont {Bruggeman}\ \emph {et~al.}(2020)\citenamefont
  {Bruggeman}, \citenamefont {Planqu{\'e}}, \citenamefont {Molenaar},\ and\
  \citenamefont {Teusink}}]{bruggeman2020}%
  \BibitemOpen
  \bibfield  {author} {\bibinfo {author} {\bibfnamefont {F.~J.}\ \bibnamefont
  {Bruggeman}}, \bibinfo {author} {\bibfnamefont {R.}~\bibnamefont
  {Planqu{\'e}}}, \bibinfo {author} {\bibfnamefont {D.}~\bibnamefont
  {Molenaar}}, \ and\ \bibinfo {author} {\bibfnamefont {B.}~\bibnamefont
  {Teusink}},\ }\href@noop {} {\bibfield  {journal} {\bibinfo  {journal} {FEMS
  Microbiology Reviews}\ }\textbf {\bibinfo {volume} {44}},\ \bibinfo {pages}
  {821} (\bibinfo {year} {2020})}\BibitemShut {NoStop}%
\bibitem [{\citenamefont {Sauls}\ \emph {et~al.}(2016)\citenamefont {Sauls},
  \citenamefont {Li},\ and\ \citenamefont {Jun}}]{sauls2016}%
  \BibitemOpen
  \bibfield  {author} {\bibinfo {author} {\bibfnamefont {J.~T.}\ \bibnamefont
  {Sauls}}, \bibinfo {author} {\bibfnamefont {D.}~\bibnamefont {Li}}, \ and\
  \bibinfo {author} {\bibfnamefont {S.}~\bibnamefont {Jun}},\ }\href@noop {}
  {\bibfield  {journal} {\bibinfo  {journal} {Current opinion in cell biology}\
  }\textbf {\bibinfo {volume} {38}},\ \bibinfo {pages} {38} (\bibinfo {year}
  {2016})}\BibitemShut {NoStop}%
\bibitem [{\citenamefont {Willis}\ and\ \citenamefont
  {Huang}(2017)}]{willis2017}%
  \BibitemOpen
  \bibfield  {author} {\bibinfo {author} {\bibfnamefont {L.}~\bibnamefont
  {Willis}}\ and\ \bibinfo {author} {\bibfnamefont {K.~C.}\ \bibnamefont
  {Huang}},\ }\href@noop {} {\bibfield  {journal} {\bibinfo  {journal} {Nature
  Reviews Microbiology}\ }\textbf {\bibinfo {volume} {15}},\ \bibinfo {pages}
  {606} (\bibinfo {year} {2017})}\BibitemShut {NoStop}%
\bibitem [{\citenamefont {Yang}\ \emph {et~al.}(2016)\citenamefont {Yang},
  \citenamefont {Blair},\ and\ \citenamefont {Salama}}]{yang2016}%
  \BibitemOpen
  \bibfield  {author} {\bibinfo {author} {\bibfnamefont {D.~C.}\ \bibnamefont
  {Yang}}, \bibinfo {author} {\bibfnamefont {K.~M.}\ \bibnamefont {Blair}}, \
  and\ \bibinfo {author} {\bibfnamefont {N.~R.}\ \bibnamefont {Salama}},\
  }\href@noop {} {\bibfield  {journal} {\bibinfo  {journal} {Microbiology and
  Molecular Biology Reviews}\ }\textbf {\bibinfo {volume} {80}},\ \bibinfo
  {pages} {187} (\bibinfo {year} {2016})}\BibitemShut {NoStop}%
\bibitem [{\citenamefont {Harris}\ and\ \citenamefont
  {Theriot}(2018)}]{harris2018}%
  \BibitemOpen
  \bibfield  {author} {\bibinfo {author} {\bibfnamefont {L.~K.}\ \bibnamefont
  {Harris}}\ and\ \bibinfo {author} {\bibfnamefont {J.~A.}\ \bibnamefont
  {Theriot}},\ }\href@noop {} {\bibfield  {journal} {\bibinfo  {journal}
  {Trends in microbiology}\ }\textbf {\bibinfo {volume} {26}},\ \bibinfo
  {pages} {815} (\bibinfo {year} {2018})}\BibitemShut {NoStop}%
\bibitem [{\citenamefont {van Teeffelen}\ and\ \citenamefont
  {Renner}(2018)}]{van2018}%
  \BibitemOpen
  \bibfield  {author} {\bibinfo {author} {\bibfnamefont {S.}~\bibnamefont {van
  Teeffelen}}\ and\ \bibinfo {author} {\bibfnamefont {L.~D.}\ \bibnamefont
  {Renner}},\ }\href@noop {} {\bibfield  {journal} {\bibinfo  {journal}
  {F1000Research}\ }\textbf {\bibinfo {volume} {7}} (\bibinfo {year}
  {2018})}\BibitemShut {NoStop}%
\bibitem [{\citenamefont {Monod}(1949)}]{monod1949}%
  \BibitemOpen
  \bibfield  {author} {\bibinfo {author} {\bibfnamefont {J.}~\bibnamefont
  {Monod}},\ }\href@noop {} {\bibfield  {journal} {\bibinfo  {journal} {Annual
  review of microbiology}\ }\textbf {\bibinfo {volume} {3}},\ \bibinfo {pages}
  {371} (\bibinfo {year} {1949})}\BibitemShut {NoStop}%
\bibitem [{\citenamefont {Belliveau}\ \emph {et~al.}(2021)\citenamefont
  {Belliveau}, \citenamefont {Chure}, \citenamefont {Hueschen}, \citenamefont
  {Garcia}, \citenamefont {Kondev}, \citenamefont {Fisher}, \citenamefont
  {Theriot},\ and\ \citenamefont {Phillips}}]{belliveau2021}%
  \BibitemOpen
  \bibfield  {author} {\bibinfo {author} {\bibfnamefont {N.~M.}\ \bibnamefont
  {Belliveau}}, \bibinfo {author} {\bibfnamefont {G.}~\bibnamefont {Chure}},
  \bibinfo {author} {\bibfnamefont {C.~L.}\ \bibnamefont {Hueschen}}, \bibinfo
  {author} {\bibfnamefont {H.~G.}\ \bibnamefont {Garcia}}, \bibinfo {author}
  {\bibfnamefont {J.}~\bibnamefont {Kondev}}, \bibinfo {author} {\bibfnamefont
  {D.~S.}\ \bibnamefont {Fisher}}, \bibinfo {author} {\bibfnamefont {J.~A.}\
  \bibnamefont {Theriot}}, \ and\ \bibinfo {author} {\bibfnamefont
  {R.}~\bibnamefont {Phillips}},\ }\href@noop {} {\bibfield  {journal}
  {\bibinfo  {journal} {Cell Systems}\ } (\bibinfo {year} {2021})}\BibitemShut
  {NoStop}%
\bibitem [{\citenamefont {Bremer}\ and\ \citenamefont
  {Dennis}(1996)}]{bremer1996}%
  \BibitemOpen
  \bibfield  {author} {\bibinfo {author} {\bibfnamefont {H.}~\bibnamefont
  {Bremer}}\ and\ \bibinfo {author} {\bibfnamefont {P.~P.}\ \bibnamefont
  {Dennis}},\ }\href@noop {} {\bibfield  {journal} {\bibinfo  {journal}
  {Washington (DC): American Society for Microbiology. Chapter, Modulation of
  chemical composition and other parameters of the cell by growth rate}\ ,\
  \bibinfo {pages} {1553}} (\bibinfo {year} {1996})}\BibitemShut {NoStop}%
\bibitem [{\citenamefont {Neidhardt}\ and\ \citenamefont
  {Magasanik}(1960)}]{neidhardt1960}%
  \BibitemOpen
  \bibfield  {author} {\bibinfo {author} {\bibfnamefont {F.~C.}\ \bibnamefont
  {Neidhardt}}\ and\ \bibinfo {author} {\bibfnamefont {B.}~\bibnamefont
  {Magasanik}},\ }\href@noop {} {\bibfield  {journal} {\bibinfo  {journal}
  {Biochimica et biophysica acta}\ }\textbf {\bibinfo {volume} {42}},\ \bibinfo
  {pages} {99} (\bibinfo {year} {1960})}\BibitemShut {NoStop}%
\bibitem [{\citenamefont {Harvey}(1973)}]{harvey1973}%
  \BibitemOpen
  \bibfield  {author} {\bibinfo {author} {\bibfnamefont {R.}~\bibnamefont
  {Harvey}},\ }\href@noop {} {\bibfield  {journal} {\bibinfo  {journal}
  {Journal of Bacteriology}\ }\textbf {\bibinfo {volume} {114}},\ \bibinfo
  {pages} {287} (\bibinfo {year} {1973})}\BibitemShut {NoStop}%
\bibitem [{\citenamefont {Scott}\ \emph {et~al.}(2010)\citenamefont {Scott},
  \citenamefont {Gunderson}, \citenamefont {Mateescu}, \citenamefont {Zhang},\
  and\ \citenamefont {Hwa}}]{scott2010}%
  \BibitemOpen
  \bibfield  {author} {\bibinfo {author} {\bibfnamefont {M.}~\bibnamefont
  {Scott}}, \bibinfo {author} {\bibfnamefont {C.~W.}\ \bibnamefont
  {Gunderson}}, \bibinfo {author} {\bibfnamefont {E.~M.}\ \bibnamefont
  {Mateescu}}, \bibinfo {author} {\bibfnamefont {Z.}~\bibnamefont {Zhang}}, \
  and\ \bibinfo {author} {\bibfnamefont {T.}~\bibnamefont {Hwa}},\ }\href@noop
  {} {\bibfield  {journal} {\bibinfo  {journal} {Science}\ }\textbf {\bibinfo
  {volume} {330}},\ \bibinfo {pages} {1099} (\bibinfo {year}
  {2010})}\BibitemShut {NoStop}%
\bibitem [{\citenamefont {Metzl-Raz}\ \emph {et~al.}(2017)\citenamefont
  {Metzl-Raz}, \citenamefont {Kafri}, \citenamefont {Yaakov}, \citenamefont
  {Soifer}, \citenamefont {Gurvich},\ and\ \citenamefont {Barkai}}]{metzl2017}%
  \BibitemOpen
  \bibfield  {author} {\bibinfo {author} {\bibfnamefont {E.}~\bibnamefont
  {Metzl-Raz}}, \bibinfo {author} {\bibfnamefont {M.}~\bibnamefont {Kafri}},
  \bibinfo {author} {\bibfnamefont {G.}~\bibnamefont {Yaakov}}, \bibinfo
  {author} {\bibfnamefont {I.}~\bibnamefont {Soifer}}, \bibinfo {author}
  {\bibfnamefont {Y.}~\bibnamefont {Gurvich}}, \ and\ \bibinfo {author}
  {\bibfnamefont {N.}~\bibnamefont {Barkai}},\ }\href@noop {} {\bibfield
  {journal} {\bibinfo  {journal} {eLife}\ }\textbf {\bibinfo {volume} {6}},\
  \bibinfo {pages} {e28034} (\bibinfo {year} {2017})}\BibitemShut {NoStop}%
\bibitem [{\citenamefont {Forchhammer}\ and\ \citenamefont
  {Lindahl}(1971)}]{forchhammer1971}%
  \BibitemOpen
  \bibfield  {author} {\bibinfo {author} {\bibfnamefont {J.}~\bibnamefont
  {Forchhammer}}\ and\ \bibinfo {author} {\bibfnamefont {L.}~\bibnamefont
  {Lindahl}},\ }\href@noop {} {\bibfield  {journal} {\bibinfo  {journal}
  {Journal of Molecular Biology}\ }\textbf {\bibinfo {volume} {55}},\ \bibinfo
  {pages} {563} (\bibinfo {year} {1971})}\BibitemShut {NoStop}%
\bibitem [{\citenamefont {Schmidt}\ \emph {et~al.}(2016)\citenamefont
  {Schmidt}, \citenamefont {Kochanowski}, \citenamefont {Vedelaar},
  \citenamefont {Ahrn{\'e}}, \citenamefont {Volkmer}, \citenamefont {Callipo},
  \citenamefont {Knoops}, \citenamefont {Bauer}, \citenamefont {Aebersold},\
  and\ \citenamefont {Heinemann}}]{schmidt2016}%
  \BibitemOpen
  \bibfield  {author} {\bibinfo {author} {\bibfnamefont {A.}~\bibnamefont
  {Schmidt}}, \bibinfo {author} {\bibfnamefont {K.}~\bibnamefont
  {Kochanowski}}, \bibinfo {author} {\bibfnamefont {S.}~\bibnamefont
  {Vedelaar}}, \bibinfo {author} {\bibfnamefont {E.}~\bibnamefont {Ahrn{\'e}}},
  \bibinfo {author} {\bibfnamefont {B.}~\bibnamefont {Volkmer}}, \bibinfo
  {author} {\bibfnamefont {L.}~\bibnamefont {Callipo}}, \bibinfo {author}
  {\bibfnamefont {K.}~\bibnamefont {Knoops}}, \bibinfo {author} {\bibfnamefont
  {M.}~\bibnamefont {Bauer}}, \bibinfo {author} {\bibfnamefont
  {R.}~\bibnamefont {Aebersold}}, \ and\ \bibinfo {author} {\bibfnamefont
  {M.}~\bibnamefont {Heinemann}},\ }\href@noop {} {\bibfield  {journal}
  {\bibinfo  {journal} {Nature Biotechnology}\ }\textbf {\bibinfo {volume}
  {34}},\ \bibinfo {pages} {104} (\bibinfo {year} {2016})}\BibitemShut
  {NoStop}%
\bibitem [{\citenamefont {Molenaar}\ \emph {et~al.}(2009)\citenamefont
  {Molenaar}, \citenamefont {Van~Berlo}, \citenamefont {De~Ridder},\ and\
  \citenamefont {Teusink}}]{molenaar2009}%
  \BibitemOpen
  \bibfield  {author} {\bibinfo {author} {\bibfnamefont {D.}~\bibnamefont
  {Molenaar}}, \bibinfo {author} {\bibfnamefont {R.}~\bibnamefont {Van~Berlo}},
  \bibinfo {author} {\bibfnamefont {D.}~\bibnamefont {De~Ridder}}, \ and\
  \bibinfo {author} {\bibfnamefont {B.}~\bibnamefont {Teusink}},\ }\href@noop
  {} {\bibfield  {journal} {\bibinfo  {journal} {Molecular Systems Biology}\
  }\textbf {\bibinfo {volume} {5}},\ \bibinfo {pages} {323} (\bibinfo {year}
  {2009})}\BibitemShut {NoStop}%
\bibitem [{\citenamefont {Klumpp}\ \emph {et~al.}(2013)\citenamefont {Klumpp},
  \citenamefont {Scott}, \citenamefont {Pedersen},\ and\ \citenamefont
  {Hwa}}]{klumpp2013}%
  \BibitemOpen
  \bibfield  {author} {\bibinfo {author} {\bibfnamefont {S.}~\bibnamefont
  {Klumpp}}, \bibinfo {author} {\bibfnamefont {M.}~\bibnamefont {Scott}},
  \bibinfo {author} {\bibfnamefont {S.}~\bibnamefont {Pedersen}}, \ and\
  \bibinfo {author} {\bibfnamefont {T.}~\bibnamefont {Hwa}},\ }\href@noop {}
  {\bibfield  {journal} {\bibinfo  {journal} {Proceedings of the National
  Academy of Sciences}\ }\textbf {\bibinfo {volume} {110}},\ \bibinfo {pages}
  {16754} (\bibinfo {year} {2013})}\BibitemShut {NoStop}%
\bibitem [{\citenamefont {Neidhardt}\ \emph {et~al.}(1990)\citenamefont
  {Neidhardt}, \citenamefont {Ingraham},\ and\ \citenamefont
  {Schaechter}}]{neidhardt1990}%
  \BibitemOpen
  \bibfield  {author} {\bibinfo {author} {\bibfnamefont {F.~C.}\ \bibnamefont
  {Neidhardt}}, \bibinfo {author} {\bibfnamefont {J.~L.}\ \bibnamefont
  {Ingraham}}, \ and\ \bibinfo {author} {\bibfnamefont {M.}~\bibnamefont
  {Schaechter}},\ }\href@noop {} {\emph {\bibinfo {title} {Physiology of the
  bacterial cell; a molecular approach}}},\ \bibinfo {number} {589.901 N397}\
  (\bibinfo  {publisher} {Sinauer associates},\ \bibinfo {year}
  {1990})\BibitemShut {NoStop}%
\bibitem [{\citenamefont {Ni}\ \emph {et~al.}(2020)\citenamefont {Ni},
  \citenamefont {Colin}, \citenamefont {Link}, \citenamefont {Endres},\ and\
  \citenamefont {Sourjik}}]{ni2020}%
  \BibitemOpen
  \bibfield  {author} {\bibinfo {author} {\bibfnamefont {B.}~\bibnamefont
  {Ni}}, \bibinfo {author} {\bibfnamefont {R.}~\bibnamefont {Colin}}, \bibinfo
  {author} {\bibfnamefont {H.}~\bibnamefont {Link}}, \bibinfo {author}
  {\bibfnamefont {R.~G.}\ \bibnamefont {Endres}}, \ and\ \bibinfo {author}
  {\bibfnamefont {V.}~\bibnamefont {Sourjik}},\ }\href@noop {} {\bibfield
  {journal} {\bibinfo  {journal} {Proceedings of the National Academy of
  Sciences}\ }\textbf {\bibinfo {volume} {117}},\ \bibinfo {pages} {595}
  (\bibinfo {year} {2020})}\BibitemShut {NoStop}%
\bibitem [{\citenamefont {Bosdriesz}\ \emph {et~al.}(2015)\citenamefont
  {Bosdriesz}, \citenamefont {Molenaar}, \citenamefont {Teusink},\ and\
  \citenamefont {Bruggeman}}]{bosdriesz2015}%
  \BibitemOpen
  \bibfield  {author} {\bibinfo {author} {\bibfnamefont {E.}~\bibnamefont
  {Bosdriesz}}, \bibinfo {author} {\bibfnamefont {D.}~\bibnamefont {Molenaar}},
  \bibinfo {author} {\bibfnamefont {B.}~\bibnamefont {Teusink}}, \ and\
  \bibinfo {author} {\bibfnamefont {F.~J.}\ \bibnamefont {Bruggeman}},\
  }\href@noop {} {\bibfield  {journal} {\bibinfo  {journal} {The FEBS Journal}\
  }\textbf {\bibinfo {volume} {282}},\ \bibinfo {pages} {2029} (\bibinfo {year}
  {2015})}\BibitemShut {NoStop}%
\bibitem [{\citenamefont {Maitra}\ and\ \citenamefont
  {Dill}(2015)}]{maitra2015}%
  \BibitemOpen
  \bibfield  {author} {\bibinfo {author} {\bibfnamefont {A.}~\bibnamefont
  {Maitra}}\ and\ \bibinfo {author} {\bibfnamefont {K.~A.}\ \bibnamefont
  {Dill}},\ }\href@noop {} {\bibfield  {journal} {\bibinfo  {journal}
  {Proceedings of the National Academy of Sciences}\ }\textbf {\bibinfo
  {volume} {112}},\ \bibinfo {pages} {406} (\bibinfo {year}
  {2015})}\BibitemShut {NoStop}%
\bibitem [{\citenamefont {Wei{\ss}e}\ \emph {et~al.}(2015)\citenamefont
  {Wei{\ss}e}, \citenamefont {Oyarz{\'u}n}, \citenamefont {Danos},\ and\
  \citenamefont {Swain}}]{weisse2015}%
  \BibitemOpen
  \bibfield  {author} {\bibinfo {author} {\bibfnamefont {A.~Y.}\ \bibnamefont
  {Wei{\ss}e}}, \bibinfo {author} {\bibfnamefont {D.~A.}\ \bibnamefont
  {Oyarz{\'u}n}}, \bibinfo {author} {\bibfnamefont {V.}~\bibnamefont {Danos}},
  \ and\ \bibinfo {author} {\bibfnamefont {P.~S.}\ \bibnamefont {Swain}},\
  }\href@noop {} {\bibfield  {journal} {\bibinfo  {journal} {Proceedings of the
  National Academy of Sciences}\ }\textbf {\bibinfo {volume} {112}},\ \bibinfo
  {pages} {E1038} (\bibinfo {year} {2015})}\BibitemShut {NoStop}%
\bibitem [{\citenamefont {Kohanim}\ \emph {et~al.}(2018)\citenamefont
  {Kohanim}, \citenamefont {Levi}, \citenamefont {Jona}, \citenamefont
  {Towbin}, \citenamefont {Bren},\ and\ \citenamefont {Alon}}]{kohanim2018}%
  \BibitemOpen
  \bibfield  {author} {\bibinfo {author} {\bibfnamefont {Y.~K.}\ \bibnamefont
  {Kohanim}}, \bibinfo {author} {\bibfnamefont {D.}~\bibnamefont {Levi}},
  \bibinfo {author} {\bibfnamefont {G.}~\bibnamefont {Jona}}, \bibinfo {author}
  {\bibfnamefont {B.~D.}\ \bibnamefont {Towbin}}, \bibinfo {author}
  {\bibfnamefont {A.}~\bibnamefont {Bren}}, \ and\ \bibinfo {author}
  {\bibfnamefont {U.}~\bibnamefont {Alon}},\ }\href@noop {} {\bibfield
  {journal} {\bibinfo  {journal} {Cell Reports}\ }\textbf {\bibinfo {volume}
  {23}},\ \bibinfo {pages} {2891} (\bibinfo {year} {2018})}\BibitemShut
  {NoStop}%
\bibitem [{\citenamefont {Pandey}\ and\ \citenamefont
  {Jain}(2016)}]{pandey2016}%
  \BibitemOpen
  \bibfield  {author} {\bibinfo {author} {\bibfnamefont {P.~P.}\ \bibnamefont
  {Pandey}}\ and\ \bibinfo {author} {\bibfnamefont {S.}~\bibnamefont {Jain}},\
  }\href@noop {} {\bibfield  {journal} {\bibinfo  {journal} {Theory in
  Biosciences}\ }\textbf {\bibinfo {volume} {135}},\ \bibinfo {pages} {121}
  (\bibinfo {year} {2016})}\BibitemShut {NoStop}%
\bibitem [{\citenamefont {Wright}\ \emph {et~al.}(2015)\citenamefont {Wright},
  \citenamefont {Banerjee}, \citenamefont {Iyer-Biswas}, \citenamefont
  {Crosson}, \citenamefont {Dinner},\ and\ \citenamefont
  {Scherer}}]{wright2015}%
  \BibitemOpen
  \bibfield  {author} {\bibinfo {author} {\bibfnamefont {C.~S.}\ \bibnamefont
  {Wright}}, \bibinfo {author} {\bibfnamefont {S.}~\bibnamefont {Banerjee}},
  \bibinfo {author} {\bibfnamefont {S.}~\bibnamefont {Iyer-Biswas}}, \bibinfo
  {author} {\bibfnamefont {S.}~\bibnamefont {Crosson}}, \bibinfo {author}
  {\bibfnamefont {A.~R.}\ \bibnamefont {Dinner}}, \ and\ \bibinfo {author}
  {\bibfnamefont {N.~F.}\ \bibnamefont {Scherer}},\ }\href@noop {} {\bibfield
  {journal} {\bibinfo  {journal} {Scientific Reports}\ }\textbf {\bibinfo
  {volume} {5}},\ \bibinfo {pages} {1} (\bibinfo {year} {2015})}\BibitemShut
  {NoStop}%
\bibitem [{\citenamefont {Wang}\ \emph {et~al.}(2010)\citenamefont {Wang},
  \citenamefont {Robert}, \citenamefont {Pelletier}, \citenamefont {Dang},
  \citenamefont {Taddei}, \citenamefont {Wright},\ and\ \citenamefont
  {Jun}}]{wang2010}%
  \BibitemOpen
  \bibfield  {author} {\bibinfo {author} {\bibfnamefont {P.}~\bibnamefont
  {Wang}}, \bibinfo {author} {\bibfnamefont {L.}~\bibnamefont {Robert}},
  \bibinfo {author} {\bibfnamefont {J.}~\bibnamefont {Pelletier}}, \bibinfo
  {author} {\bibfnamefont {W.~L.}\ \bibnamefont {Dang}}, \bibinfo {author}
  {\bibfnamefont {F.}~\bibnamefont {Taddei}}, \bibinfo {author} {\bibfnamefont
  {A.}~\bibnamefont {Wright}}, \ and\ \bibinfo {author} {\bibfnamefont
  {S.}~\bibnamefont {Jun}},\ }\href@noop {} {\bibfield  {journal} {\bibinfo
  {journal} {Current Biology}\ }\textbf {\bibinfo {volume} {20}},\ \bibinfo
  {pages} {1099} (\bibinfo {year} {2010})}\BibitemShut {NoStop}%
\bibitem [{\citenamefont {Cadart}\ \emph {et~al.}(2018)\citenamefont {Cadart},
  \citenamefont {Monnier}, \citenamefont {Grilli}, \citenamefont {S{\'a}ez},
  \citenamefont {Srivastava}, \citenamefont {Attia}, \citenamefont {Terriac},
  \citenamefont {Baum}, \citenamefont {Cosentino-Lagomarsino},\ and\
  \citenamefont {Piel}}]{cadart2018}%
  \BibitemOpen
  \bibfield  {author} {\bibinfo {author} {\bibfnamefont {C.}~\bibnamefont
  {Cadart}}, \bibinfo {author} {\bibfnamefont {S.}~\bibnamefont {Monnier}},
  \bibinfo {author} {\bibfnamefont {J.}~\bibnamefont {Grilli}}, \bibinfo
  {author} {\bibfnamefont {P.~J.}\ \bibnamefont {S{\'a}ez}}, \bibinfo {author}
  {\bibfnamefont {N.}~\bibnamefont {Srivastava}}, \bibinfo {author}
  {\bibfnamefont {R.}~\bibnamefont {Attia}}, \bibinfo {author} {\bibfnamefont
  {E.}~\bibnamefont {Terriac}}, \bibinfo {author} {\bibfnamefont
  {B.}~\bibnamefont {Baum}}, \bibinfo {author} {\bibfnamefont {M.}~\bibnamefont
  {Cosentino-Lagomarsino}}, \ and\ \bibinfo {author} {\bibfnamefont
  {M.}~\bibnamefont {Piel}},\ }\href@noop {} {\bibfield  {journal} {\bibinfo
  {journal} {Nature Communications}\ }\textbf {\bibinfo {volume} {9}},\
  \bibinfo {pages} {1} (\bibinfo {year} {2018})}\BibitemShut {NoStop}%
\bibitem [{\citenamefont {Zheng}\ \emph {et~al.}(2020)\citenamefont {Zheng},
  \citenamefont {Bai}, \citenamefont {Jiang}, \citenamefont {Tokuyasu},
  \citenamefont {Huang}, \citenamefont {Zhong}, \citenamefont {Wu},
  \citenamefont {Fu}, \citenamefont {Kleckner}, \citenamefont {Hwa} \emph
  {et~al.}}]{zheng2020}%
  \BibitemOpen
  \bibfield  {author} {\bibinfo {author} {\bibfnamefont {H.}~\bibnamefont
  {Zheng}}, \bibinfo {author} {\bibfnamefont {Y.}~\bibnamefont {Bai}}, \bibinfo
  {author} {\bibfnamefont {M.}~\bibnamefont {Jiang}}, \bibinfo {author}
  {\bibfnamefont {T.~A.}\ \bibnamefont {Tokuyasu}}, \bibinfo {author}
  {\bibfnamefont {X.}~\bibnamefont {Huang}}, \bibinfo {author} {\bibfnamefont
  {F.}~\bibnamefont {Zhong}}, \bibinfo {author} {\bibfnamefont
  {Y.}~\bibnamefont {Wu}}, \bibinfo {author} {\bibfnamefont {X.}~\bibnamefont
  {Fu}}, \bibinfo {author} {\bibfnamefont {N.}~\bibnamefont {Kleckner}},
  \bibinfo {author} {\bibfnamefont {T.}~\bibnamefont {Hwa}},  \emph {et~al.},\
  }\href@noop {} {\bibfield  {journal} {\bibinfo  {journal} {Nature
  Microbiology}\ }\textbf {\bibinfo {volume} {5}},\ \bibinfo {pages} {995}
  (\bibinfo {year} {2020})}\BibitemShut {NoStop}%
\bibitem [{\citenamefont {Zhu}\ \emph {et~al.}(2017)\citenamefont {Zhu},
  \citenamefont {Dai}, \citenamefont {Guo}, \citenamefont {Ge}, \citenamefont
  {Yang}, \citenamefont {Wang},\ and\ \citenamefont {Wang}}]{zhu2017}%
  \BibitemOpen
  \bibfield  {author} {\bibinfo {author} {\bibfnamefont {M.}~\bibnamefont
  {Zhu}}, \bibinfo {author} {\bibfnamefont {X.}~\bibnamefont {Dai}}, \bibinfo
  {author} {\bibfnamefont {W.}~\bibnamefont {Guo}}, \bibinfo {author}
  {\bibfnamefont {Z.}~\bibnamefont {Ge}}, \bibinfo {author} {\bibfnamefont
  {M.}~\bibnamefont {Yang}}, \bibinfo {author} {\bibfnamefont {H.}~\bibnamefont
  {Wang}}, \ and\ \bibinfo {author} {\bibfnamefont {Y.-P.}\ \bibnamefont
  {Wang}},\ }\href@noop {} {\bibfield  {journal} {\bibinfo  {journal} {MBio}\
  }\textbf {\bibinfo {volume} {8}} (\bibinfo {year} {2017})}\BibitemShut
  {NoStop}%
\bibitem [{\citenamefont {Bertaux}\ \emph {et~al.}(2020)\citenamefont
  {Bertaux}, \citenamefont {Von~K{\"u}gelgen}, \citenamefont {Marguerat},\ and\
  \citenamefont {Shahrezaei}}]{bertaux2020}%
  \BibitemOpen
  \bibfield  {author} {\bibinfo {author} {\bibfnamefont {F.}~\bibnamefont
  {Bertaux}}, \bibinfo {author} {\bibfnamefont {J.}~\bibnamefont
  {Von~K{\"u}gelgen}}, \bibinfo {author} {\bibfnamefont {S.}~\bibnamefont
  {Marguerat}}, \ and\ \bibinfo {author} {\bibfnamefont {V.}~\bibnamefont
  {Shahrezaei}},\ }\href@noop {} {\bibfield  {journal} {\bibinfo  {journal}
  {PLoS computational biology}\ }\textbf {\bibinfo {volume} {16}},\ \bibinfo
  {pages} {e1008245} (\bibinfo {year} {2020})}\BibitemShut {NoStop}%
\bibitem [{\citenamefont {Mori}\ \emph {et~al.}(2021)\citenamefont {Mori},
  \citenamefont {Zhang}, \citenamefont {Banaei-Esfahani}, \citenamefont
  {Lalanne}, \citenamefont {Okano}, \citenamefont {Collins}, \citenamefont
  {Schmidt}, \citenamefont {Schubert}, \citenamefont {Lee}, \citenamefont {Li}
  \emph {et~al.}}]{mori2021}%
  \BibitemOpen
  \bibfield  {author} {\bibinfo {author} {\bibfnamefont {M.}~\bibnamefont
  {Mori}}, \bibinfo {author} {\bibfnamefont {Z.}~\bibnamefont {Zhang}},
  \bibinfo {author} {\bibfnamefont {A.}~\bibnamefont {Banaei-Esfahani}},
  \bibinfo {author} {\bibfnamefont {J.-B.}\ \bibnamefont {Lalanne}}, \bibinfo
  {author} {\bibfnamefont {H.}~\bibnamefont {Okano}}, \bibinfo {author}
  {\bibfnamefont {B.~C.}\ \bibnamefont {Collins}}, \bibinfo {author}
  {\bibfnamefont {A.}~\bibnamefont {Schmidt}}, \bibinfo {author} {\bibfnamefont
  {O.~T.}\ \bibnamefont {Schubert}}, \bibinfo {author} {\bibfnamefont {D.-S.}\
  \bibnamefont {Lee}}, \bibinfo {author} {\bibfnamefont {G.-W.}\ \bibnamefont
  {Li}},  \emph {et~al.},\ }\href@noop {} {\bibfield  {journal} {\bibinfo
  {journal} {Molecular Systems Biology}\ }\textbf {\bibinfo {volume} {17}},\
  \bibinfo {pages} {e9536} (\bibinfo {year} {2021})}\BibitemShut {NoStop}%
\bibitem [{\citenamefont {Adams}\ and\ \citenamefont
  {Errington}(2009)}]{adams2009}%
  \BibitemOpen
  \bibfield  {author} {\bibinfo {author} {\bibfnamefont {D.~W.}\ \bibnamefont
  {Adams}}\ and\ \bibinfo {author} {\bibfnamefont {J.}~\bibnamefont
  {Errington}},\ }\href@noop {} {\bibfield  {journal} {\bibinfo  {journal}
  {Nature Reviews Microbiology}\ }\textbf {\bibinfo {volume} {7}},\ \bibinfo
  {pages} {642} (\bibinfo {year} {2009})}\BibitemShut {NoStop}%
\bibitem [{\citenamefont {Bi}\ and\ \citenamefont {Lutkenhaus}(1991)}]{bi1991}%
  \BibitemOpen
  \bibfield  {author} {\bibinfo {author} {\bibfnamefont {E.}~\bibnamefont
  {Bi}}\ and\ \bibinfo {author} {\bibfnamefont {J.}~\bibnamefont
  {Lutkenhaus}},\ }\href@noop {} {\bibfield  {journal} {\bibinfo  {journal}
  {Nature}\ }\textbf {\bibinfo {volume} {354}},\ \bibinfo {pages} {161}
  (\bibinfo {year} {1991})}\BibitemShut {NoStop}%
\bibitem [{\citenamefont {Fl{\aa}tten}\ \emph {et~al.}(2015)\citenamefont
  {Fl{\aa}tten}, \citenamefont {Fossum-Raunehaug}, \citenamefont {Taipale},
  \citenamefont {Martinsen},\ and\ \citenamefont {Skarstad}}]{flaatten2015}%
  \BibitemOpen
  \bibfield  {author} {\bibinfo {author} {\bibfnamefont {I.}~\bibnamefont
  {Fl{\aa}tten}}, \bibinfo {author} {\bibfnamefont {S.}~\bibnamefont
  {Fossum-Raunehaug}}, \bibinfo {author} {\bibfnamefont {R.}~\bibnamefont
  {Taipale}}, \bibinfo {author} {\bibfnamefont {S.}~\bibnamefont {Martinsen}},
  \ and\ \bibinfo {author} {\bibfnamefont {K.}~\bibnamefont {Skarstad}},\
  }\href@noop {} {\bibfield  {journal} {\bibinfo  {journal} {PLoS Genetics}\
  }\textbf {\bibinfo {volume} {11}},\ \bibinfo {pages} {e1005276} (\bibinfo
  {year} {2015})}\BibitemShut {NoStop}%
\bibitem [{\citenamefont {Ojkic}\ and\ \citenamefont
  {Banerjee}(2021)}]{ojkic2021}%
  \BibitemOpen
  \bibfield  {author} {\bibinfo {author} {\bibfnamefont {N.}~\bibnamefont
  {Ojkic}}\ and\ \bibinfo {author} {\bibfnamefont {S.}~\bibnamefont
  {Banerjee}},\ }\href@noop {} {\bibfield  {journal} {\bibinfo  {journal}
  {Biophysical Journal}\ } (\bibinfo {year} {2021})}\BibitemShut {NoStop}%
\bibitem [{\citenamefont {Weart}\ \emph {et~al.}(2007)\citenamefont {Weart},
  \citenamefont {Lee}, \citenamefont {Chien}, \citenamefont {Haeusser},
  \citenamefont {Hill},\ and\ \citenamefont {Levin}}]{weart2007}%
  \BibitemOpen
  \bibfield  {author} {\bibinfo {author} {\bibfnamefont {R.~B.}\ \bibnamefont
  {Weart}}, \bibinfo {author} {\bibfnamefont {A.~H.}\ \bibnamefont {Lee}},
  \bibinfo {author} {\bibfnamefont {A.-C.}\ \bibnamefont {Chien}}, \bibinfo
  {author} {\bibfnamefont {D.~P.}\ \bibnamefont {Haeusser}}, \bibinfo {author}
  {\bibfnamefont {N.~S.}\ \bibnamefont {Hill}}, \ and\ \bibinfo {author}
  {\bibfnamefont {P.~A.}\ \bibnamefont {Levin}},\ }\href@noop {} {\bibfield
  {journal} {\bibinfo  {journal} {Cell}\ }\textbf {\bibinfo {volume} {130}},\
  \bibinfo {pages} {335} (\bibinfo {year} {2007})}\BibitemShut {NoStop}%
\bibitem [{\citenamefont {Chien}\ \emph {et~al.}(2012)\citenamefont {Chien},
  \citenamefont {Zareh}, \citenamefont {Wang},\ and\ \citenamefont
  {Levin}}]{chien2012}%
  \BibitemOpen
  \bibfield  {author} {\bibinfo {author} {\bibfnamefont {A.-C.}\ \bibnamefont
  {Chien}}, \bibinfo {author} {\bibfnamefont {S.~K.~G.}\ \bibnamefont {Zareh}},
  \bibinfo {author} {\bibfnamefont {Y.~M.}\ \bibnamefont {Wang}}, \ and\
  \bibinfo {author} {\bibfnamefont {P.~A.}\ \bibnamefont {Levin}},\ }\href@noop
  {} {\bibfield  {journal} {\bibinfo  {journal} {Molecular microbiology}\
  }\textbf {\bibinfo {volume} {86}},\ \bibinfo {pages} {594} (\bibinfo {year}
  {2012})}\BibitemShut {NoStop}%
\bibitem [{\citenamefont {Hill}\ \emph {et~al.}(2018)\citenamefont {Hill},
  \citenamefont {Zuke}, \citenamefont {Buske}, \citenamefont {Chien},\ and\
  \citenamefont {Levin}}]{hill2018}%
  \BibitemOpen
  \bibfield  {author} {\bibinfo {author} {\bibfnamefont {N.~S.}\ \bibnamefont
  {Hill}}, \bibinfo {author} {\bibfnamefont {J.~D.}\ \bibnamefont {Zuke}},
  \bibinfo {author} {\bibfnamefont {P.~J.}\ \bibnamefont {Buske}}, \bibinfo
  {author} {\bibfnamefont {A.-C.}\ \bibnamefont {Chien}}, \ and\ \bibinfo
  {author} {\bibfnamefont {P.~A.}\ \bibnamefont {Levin}},\ }\href@noop {}
  {\bibfield  {journal} {\bibinfo  {journal} {BMC microbiology}\ }\textbf
  {\bibinfo {volume} {18}},\ \bibinfo {pages} {1} (\bibinfo {year}
  {2018})}\BibitemShut {NoStop}%
\bibitem [{\citenamefont {Sharpe}\ \emph {et~al.}(1998)\citenamefont {Sharpe},
  \citenamefont {Hauser}, \citenamefont {Sharpe},\ and\ \citenamefont
  {Errington}}]{sharpe1998}%
  \BibitemOpen
  \bibfield  {author} {\bibinfo {author} {\bibfnamefont {M.~E.}\ \bibnamefont
  {Sharpe}}, \bibinfo {author} {\bibfnamefont {P.~M.}\ \bibnamefont {Hauser}},
  \bibinfo {author} {\bibfnamefont {R.~G.}\ \bibnamefont {Sharpe}}, \ and\
  \bibinfo {author} {\bibfnamefont {J.}~\bibnamefont {Errington}},\ }\href@noop
  {} {\bibfield  {journal} {\bibinfo  {journal} {Journal of bacteriology}\
  }\textbf {\bibinfo {volume} {180}},\ \bibinfo {pages} {547} (\bibinfo {year}
  {1998})}\BibitemShut {NoStop}%
\bibitem [{\citenamefont {Cooper}\ and\ \citenamefont
  {Helmstetter}(1968)}]{cooper1968}%
  \BibitemOpen
  \bibfield  {author} {\bibinfo {author} {\bibfnamefont {S.}~\bibnamefont
  {Cooper}}\ and\ \bibinfo {author} {\bibfnamefont {C.~E.}\ \bibnamefont
  {Helmstetter}},\ }\href@noop {} {\bibfield  {journal} {\bibinfo  {journal}
  {Journal of molecular biology}\ }\textbf {\bibinfo {volume} {31}},\ \bibinfo
  {pages} {519} (\bibinfo {year} {1968})}\BibitemShut {NoStop}%
\bibitem [{\citenamefont {Cooper}(1997)}]{cooper1997}%
  \BibitemOpen
  \bibfield  {author} {\bibinfo {author} {\bibfnamefont {S.}~\bibnamefont
  {Cooper}},\ }\href@noop {} {\bibfield  {journal} {\bibinfo  {journal}
  {Molecular microbiology}\ }\textbf {\bibinfo {volume} {26}},\ \bibinfo
  {pages} {1138} (\bibinfo {year} {1997})}\BibitemShut {NoStop}%
\bibitem [{\citenamefont {Micali}\ \emph
  {et~al.}(2018{\natexlab{a}})\citenamefont {Micali}, \citenamefont {Grilli},
  \citenamefont {Marchi}, \citenamefont {Osella},\ and\ \citenamefont
  {Lagomarsino}}]{micali2018a}%
  \BibitemOpen
  \bibfield  {author} {\bibinfo {author} {\bibfnamefont {G.}~\bibnamefont
  {Micali}}, \bibinfo {author} {\bibfnamefont {J.}~\bibnamefont {Grilli}},
  \bibinfo {author} {\bibfnamefont {J.}~\bibnamefont {Marchi}}, \bibinfo
  {author} {\bibfnamefont {M.}~\bibnamefont {Osella}}, \ and\ \bibinfo {author}
  {\bibfnamefont {M.~C.}\ \bibnamefont {Lagomarsino}},\ }\href@noop {}
  {\bibfield  {journal} {\bibinfo  {journal} {Cell reports}\ }\textbf {\bibinfo
  {volume} {25}},\ \bibinfo {pages} {761} (\bibinfo {year}
  {2018}{\natexlab{a}})}\BibitemShut {NoStop}%
\bibitem [{\citenamefont {Micali}\ \emph
  {et~al.}(2018{\natexlab{b}})\citenamefont {Micali}, \citenamefont {Grilli},
  \citenamefont {Osella},\ and\ \citenamefont {Lagomarsino}}]{micali2018b}%
  \BibitemOpen
  \bibfield  {author} {\bibinfo {author} {\bibfnamefont {G.}~\bibnamefont
  {Micali}}, \bibinfo {author} {\bibfnamefont {J.}~\bibnamefont {Grilli}},
  \bibinfo {author} {\bibfnamefont {M.}~\bibnamefont {Osella}}, \ and\ \bibinfo
  {author} {\bibfnamefont {M.~C.}\ \bibnamefont {Lagomarsino}},\ }\href@noop {}
  {\bibfield  {journal} {\bibinfo  {journal} {Science Advances}\ }\textbf
  {\bibinfo {volume} {4}},\ \bibinfo {pages} {eaau3324} (\bibinfo {year}
  {2018}{\natexlab{b}})}\BibitemShut {NoStop}%
\bibitem [{\citenamefont {Grilli}\ \emph {et~al.}(2018)\citenamefont {Grilli},
  \citenamefont {Cadart}, \citenamefont {Micali}, \citenamefont {Osella},\ and\
  \citenamefont {Cosentino~Lagomarsino}}]{grilli2018}%
  \BibitemOpen
  \bibfield  {author} {\bibinfo {author} {\bibfnamefont {J.}~\bibnamefont
  {Grilli}}, \bibinfo {author} {\bibfnamefont {C.}~\bibnamefont {Cadart}},
  \bibinfo {author} {\bibfnamefont {G.}~\bibnamefont {Micali}}, \bibinfo
  {author} {\bibfnamefont {M.}~\bibnamefont {Osella}}, \ and\ \bibinfo {author}
  {\bibfnamefont {M.}~\bibnamefont {Cosentino~Lagomarsino}},\ }\href@noop {}
  {\bibfield  {journal} {\bibinfo  {journal} {Frontiers in microbiology}\
  }\textbf {\bibinfo {volume} {9}},\ \bibinfo {pages} {1541} (\bibinfo {year}
  {2018})}\BibitemShut {NoStop}%
\bibitem [{\citenamefont {Witz}\ \emph {et~al.}(2019)\citenamefont {Witz},
  \citenamefont {van Nimwegen},\ and\ \citenamefont {Julou}}]{witz2019}%
  \BibitemOpen
  \bibfield  {author} {\bibinfo {author} {\bibfnamefont {G.}~\bibnamefont
  {Witz}}, \bibinfo {author} {\bibfnamefont {E.}~\bibnamefont {van Nimwegen}},
  \ and\ \bibinfo {author} {\bibfnamefont {T.}~\bibnamefont {Julou}},\
  }\href@noop {} {\bibfield  {journal} {\bibinfo  {journal} {Elife}\ }\textbf
  {\bibinfo {volume} {8}},\ \bibinfo {pages} {e48063} (\bibinfo {year}
  {2019})}\BibitemShut {NoStop}%
\bibitem [{\citenamefont {Le~Treut}\ \emph {et~al.}(2021)\citenamefont
  {Le~Treut}, \citenamefont {Si}, \citenamefont {Li},\ and\ \citenamefont
  {Jun}}]{le2021}%
  \BibitemOpen
  \bibfield  {author} {\bibinfo {author} {\bibfnamefont {G.}~\bibnamefont
  {Le~Treut}}, \bibinfo {author} {\bibfnamefont {F.}~\bibnamefont {Si}},
  \bibinfo {author} {\bibfnamefont {D.}~\bibnamefont {Li}}, \ and\ \bibinfo
  {author} {\bibfnamefont {S.}~\bibnamefont {Jun}},\ }\href@noop {} {\bibfield
  {journal} {\bibinfo  {journal} {bioRxiv}\ } (\bibinfo {year}
  {2021})}\BibitemShut {NoStop}%
\bibitem [{\citenamefont {McCoy}\ \emph {et~al.}(2011)\citenamefont {McCoy},
  \citenamefont {Xie},\ and\ \citenamefont {Tor}}]{mccoy2011}%
  \BibitemOpen
  \bibfield  {author} {\bibinfo {author} {\bibfnamefont {L.~S.}\ \bibnamefont
  {McCoy}}, \bibinfo {author} {\bibfnamefont {Y.}~\bibnamefont {Xie}}, \ and\
  \bibinfo {author} {\bibfnamefont {Y.}~\bibnamefont {Tor}},\ }\href@noop {}
  {\bibfield  {journal} {\bibinfo  {journal} {Wiley Interdisciplinary Reviews:
  RNA}\ }\textbf {\bibinfo {volume} {2}},\ \bibinfo {pages} {209} (\bibinfo
  {year} {2011})}\BibitemShut {NoStop}%
\bibitem [{\citenamefont {Banerjee}\ \emph {et~al.}(2021)\citenamefont
  {Banerjee}, \citenamefont {Lo}, \citenamefont {Ojkic}, \citenamefont
  {Stephens}, \citenamefont {Scherer},\ and\ \citenamefont
  {Dinner}}]{banerjee2021}%
  \BibitemOpen
  \bibfield  {author} {\bibinfo {author} {\bibfnamefont {S.}~\bibnamefont
  {Banerjee}}, \bibinfo {author} {\bibfnamefont {K.}~\bibnamefont {Lo}},
  \bibinfo {author} {\bibfnamefont {N.}~\bibnamefont {Ojkic}}, \bibinfo
  {author} {\bibfnamefont {R.}~\bibnamefont {Stephens}}, \bibinfo {author}
  {\bibfnamefont {N.~F.}\ \bibnamefont {Scherer}}, \ and\ \bibinfo {author}
  {\bibfnamefont {A.~R.}\ \bibnamefont {Dinner}},\ }\href@noop {} {\bibfield
  {journal} {\bibinfo  {journal} {Nature Physics}\ }\textbf {\bibinfo {volume}
  {17}},\ \bibinfo {pages} {403} (\bibinfo {year} {2021})}\BibitemShut
  {NoStop}%
\bibitem [{\citenamefont {Elf}\ \emph {et~al.}(2006)\citenamefont {Elf},
  \citenamefont {Nilsson}, \citenamefont {Tenson},\ and\ \citenamefont
  {Ehrenberg}}]{elf2006}%
  \BibitemOpen
  \bibfield  {author} {\bibinfo {author} {\bibfnamefont {J.}~\bibnamefont
  {Elf}}, \bibinfo {author} {\bibfnamefont {K.}~\bibnamefont {Nilsson}},
  \bibinfo {author} {\bibfnamefont {T.}~\bibnamefont {Tenson}}, \ and\ \bibinfo
  {author} {\bibfnamefont {M.}~\bibnamefont {Ehrenberg}},\ }\href@noop {}
  {\bibfield  {journal} {\bibinfo  {journal} {Physical review letters}\
  }\textbf {\bibinfo {volume} {97}},\ \bibinfo {pages} {258104} (\bibinfo
  {year} {2006})}\BibitemShut {NoStop}%
\bibitem [{\citenamefont {Greulich}\ \emph {et~al.}(2015)\citenamefont
  {Greulich}, \citenamefont {Scott}, \citenamefont {Evans},\ and\ \citenamefont
  {Allen}}]{greulich2015}%
  \BibitemOpen
  \bibfield  {author} {\bibinfo {author} {\bibfnamefont {P.}~\bibnamefont
  {Greulich}}, \bibinfo {author} {\bibfnamefont {M.}~\bibnamefont {Scott}},
  \bibinfo {author} {\bibfnamefont {M.~R.}\ \bibnamefont {Evans}}, \ and\
  \bibinfo {author} {\bibfnamefont {R.~J.}\ \bibnamefont {Allen}},\ }\href@noop
  {} {\bibfield  {journal} {\bibinfo  {journal} {Molecular systems biology}\
  }\textbf {\bibinfo {volume} {11}},\ \bibinfo {pages} {796} (\bibinfo {year}
  {2015})}\BibitemShut {NoStop}%
\bibitem [{\citenamefont {Greulich}\ \emph {et~al.}(2017)\citenamefont
  {Greulich}, \citenamefont {Dole{\v{z}}al}, \citenamefont {Scott},
  \citenamefont {Evans},\ and\ \citenamefont {Allen}}]{greulich2017}%
  \BibitemOpen
  \bibfield  {author} {\bibinfo {author} {\bibfnamefont {P.}~\bibnamefont
  {Greulich}}, \bibinfo {author} {\bibfnamefont {J.}~\bibnamefont
  {Dole{\v{z}}al}}, \bibinfo {author} {\bibfnamefont {M.}~\bibnamefont
  {Scott}}, \bibinfo {author} {\bibfnamefont {M.~R.}\ \bibnamefont {Evans}}, \
  and\ \bibinfo {author} {\bibfnamefont {R.~J.}\ \bibnamefont {Allen}},\
  }\href@noop {} {\bibfield  {journal} {\bibinfo  {journal} {Physical biology}\
  }\textbf {\bibinfo {volume} {14}},\ \bibinfo {pages} {065005} (\bibinfo
  {year} {2017})}\BibitemShut {NoStop}%
\bibitem [{\citenamefont {Ojkic}\ \emph {et~al.}(2016)\citenamefont {Ojkic},
  \citenamefont {L{\'o}pez-Garrido}, \citenamefont {Pogliano},\ and\
  \citenamefont {Endres}}]{ojkic2016}%
  \BibitemOpen
  \bibfield  {author} {\bibinfo {author} {\bibfnamefont {N.}~\bibnamefont
  {Ojkic}}, \bibinfo {author} {\bibfnamefont {J.}~\bibnamefont
  {L{\'o}pez-Garrido}}, \bibinfo {author} {\bibfnamefont {K.}~\bibnamefont
  {Pogliano}}, \ and\ \bibinfo {author} {\bibfnamefont {R.~G.}\ \bibnamefont
  {Endres}},\ }\href@noop {} {\bibfield  {journal} {\bibinfo  {journal}
  {Elife}\ }\textbf {\bibinfo {volume} {5}},\ \bibinfo {pages} {e18657}
  (\bibinfo {year} {2016})}\BibitemShut {NoStop}%
\bibitem [{\citenamefont {L{\'o}pez-Garrido}\ \emph {et~al.}(2018)\citenamefont
  {L{\'o}pez-Garrido}, \citenamefont {Ojkic}, \citenamefont {Khanna},
  \citenamefont {Wagner}, \citenamefont {Villa}, \citenamefont {Endres},\ and\
  \citenamefont {Pogliano}}]{lopez2018}%
  \BibitemOpen
  \bibfield  {author} {\bibinfo {author} {\bibfnamefont {J.}~\bibnamefont
  {L{\'o}pez-Garrido}}, \bibinfo {author} {\bibfnamefont {N.}~\bibnamefont
  {Ojkic}}, \bibinfo {author} {\bibfnamefont {K.}~\bibnamefont {Khanna}},
  \bibinfo {author} {\bibfnamefont {F.~R.}\ \bibnamefont {Wagner}}, \bibinfo
  {author} {\bibfnamefont {E.}~\bibnamefont {Villa}}, \bibinfo {author}
  {\bibfnamefont {R.~G.}\ \bibnamefont {Endres}}, \ and\ \bibinfo {author}
  {\bibfnamefont {K.}~\bibnamefont {Pogliano}},\ }\href@noop {} {\bibfield
  {journal} {\bibinfo  {journal} {Cell}\ }\textbf {\bibinfo {volume} {172}},\
  \bibinfo {pages} {758} (\bibinfo {year} {2018})}\BibitemShut {NoStop}%
\bibitem [{\citenamefont {Zaritsky}\ and\ \citenamefont
  {Pritchard}(1973)}]{zaritsky1973}%
  \BibitemOpen
  \bibfield  {author} {\bibinfo {author} {\bibfnamefont {A.}~\bibnamefont
  {Zaritsky}}\ and\ \bibinfo {author} {\bibfnamefont {R.}~\bibnamefont
  {Pritchard}},\ }\href@noop {} {\bibfield  {journal} {\bibinfo  {journal}
  {Journal of Bacteriology}\ }\textbf {\bibinfo {volume} {114}},\ \bibinfo
  {pages} {824} (\bibinfo {year} {1973})}\BibitemShut {NoStop}%
\bibitem [{\citenamefont {Zaritsky}(1975)}]{zaritsky1975}%
  \BibitemOpen
  \bibfield  {author} {\bibinfo {author} {\bibfnamefont {A.}~\bibnamefont
  {Zaritsky}},\ }\href@noop {} {\bibfield  {journal} {\bibinfo  {journal}
  {Journal of Theoretical Biology}\ }\textbf {\bibinfo {volume} {54}},\
  \bibinfo {pages} {243} (\bibinfo {year} {1975})}\BibitemShut {NoStop}%
\bibitem [{\citenamefont {Jiang}\ and\ \citenamefont {Sun}(2010)}]{jiang2010}%
  \BibitemOpen
  \bibfield  {author} {\bibinfo {author} {\bibfnamefont {H.}~\bibnamefont
  {Jiang}}\ and\ \bibinfo {author} {\bibfnamefont {S.~X.}\ \bibnamefont
  {Sun}},\ }\href@noop {} {\bibfield  {journal} {\bibinfo  {journal} {Physical
  Review Letters}\ }\textbf {\bibinfo {volume} {105}},\ \bibinfo {pages}
  {028101} (\bibinfo {year} {2010})}\BibitemShut {NoStop}%
\bibitem [{\citenamefont {Jiang}\ \emph {et~al.}(2011)\citenamefont {Jiang},
  \citenamefont {Si}, \citenamefont {Margolin},\ and\ \citenamefont
  {Sun}}]{jiang2011}%
  \BibitemOpen
  \bibfield  {author} {\bibinfo {author} {\bibfnamefont {H.}~\bibnamefont
  {Jiang}}, \bibinfo {author} {\bibfnamefont {F.}~\bibnamefont {Si}}, \bibinfo
  {author} {\bibfnamefont {W.}~\bibnamefont {Margolin}}, \ and\ \bibinfo
  {author} {\bibfnamefont {S.~X.}\ \bibnamefont {Sun}},\ }\href@noop {}
  {\bibfield  {journal} {\bibinfo  {journal} {Biophysical Journal}\ }\textbf
  {\bibinfo {volume} {101}},\ \bibinfo {pages} {327} (\bibinfo {year}
  {2011})}\BibitemShut {NoStop}%
\bibitem [{\citenamefont {Banerjee}\ \emph {et~al.}(2016)\citenamefont
  {Banerjee}, \citenamefont {Scherer},\ and\ \citenamefont
  {Dinner}}]{banerjee2016}%
  \BibitemOpen
  \bibfield  {author} {\bibinfo {author} {\bibfnamefont {S.}~\bibnamefont
  {Banerjee}}, \bibinfo {author} {\bibfnamefont {N.~F.}\ \bibnamefont
  {Scherer}}, \ and\ \bibinfo {author} {\bibfnamefont {A.~R.}\ \bibnamefont
  {Dinner}},\ }\href@noop {} {\bibfield  {journal} {\bibinfo  {journal} {Soft
  Matter}\ }\textbf {\bibinfo {volume} {12}},\ \bibinfo {pages} {3442}
  (\bibinfo {year} {2016})}\BibitemShut {NoStop}%
\bibitem [{\citenamefont {Nguyen}\ \emph {et~al.}(2020)\citenamefont {Nguyen},
  \citenamefont {Fernandez}, \citenamefont {Pontrelli}, \citenamefont {Sauer},
  \citenamefont {Ackermann},\ and\ \citenamefont {Stocker}}]{nguyen2020}%
  \BibitemOpen
  \bibfield  {author} {\bibinfo {author} {\bibfnamefont {J.}~\bibnamefont
  {Nguyen}}, \bibinfo {author} {\bibfnamefont {V.}~\bibnamefont {Fernandez}},
  \bibinfo {author} {\bibfnamefont {S.}~\bibnamefont {Pontrelli}}, \bibinfo
  {author} {\bibfnamefont {U.}~\bibnamefont {Sauer}}, \bibinfo {author}
  {\bibfnamefont {M.}~\bibnamefont {Ackermann}}, \ and\ \bibinfo {author}
  {\bibfnamefont {R.}~\bibnamefont {Stocker}},\ }\href@noop {} {\bibfield
  {journal} {\bibinfo  {journal} {BioRxiv}\ } (\bibinfo {year}
  {2020})}\BibitemShut {NoStop}%
\bibitem [{\citenamefont {Shi}\ \emph {et~al.}(2021)\citenamefont {Shi},
  \citenamefont {Hu}, \citenamefont {Odermatt}, \citenamefont {Gonzalez},
  \citenamefont {Zhang}, \citenamefont {Elias}, \citenamefont {Chang},\ and\
  \citenamefont {Huang}}]{shi2021}%
  \BibitemOpen
  \bibfield  {author} {\bibinfo {author} {\bibfnamefont {H.}~\bibnamefont
  {Shi}}, \bibinfo {author} {\bibfnamefont {Y.}~\bibnamefont {Hu}}, \bibinfo
  {author} {\bibfnamefont {P.~D.}\ \bibnamefont {Odermatt}}, \bibinfo {author}
  {\bibfnamefont {C.~G.}\ \bibnamefont {Gonzalez}}, \bibinfo {author}
  {\bibfnamefont {L.}~\bibnamefont {Zhang}}, \bibinfo {author} {\bibfnamefont
  {J.~E.}\ \bibnamefont {Elias}}, \bibinfo {author} {\bibfnamefont
  {F.}~\bibnamefont {Chang}}, \ and\ \bibinfo {author} {\bibfnamefont {K.~C.}\
  \bibnamefont {Huang}},\ }\href@noop {} {\bibfield  {journal} {\bibinfo
  {journal} {Nature communications}\ }\textbf {\bibinfo {volume} {12}},\
  \bibinfo {pages} {1} (\bibinfo {year} {2021})}\BibitemShut {NoStop}%
\bibitem [{\citenamefont {Nonejuie}\ \emph {et~al.}(2013)\citenamefont
  {Nonejuie}, \citenamefont {Burkart}, \citenamefont {Pogliano},\ and\
  \citenamefont {Pogliano}}]{nonejuie2013}%
  \BibitemOpen
  \bibfield  {author} {\bibinfo {author} {\bibfnamefont {P.}~\bibnamefont
  {Nonejuie}}, \bibinfo {author} {\bibfnamefont {M.}~\bibnamefont {Burkart}},
  \bibinfo {author} {\bibfnamefont {K.}~\bibnamefont {Pogliano}}, \ and\
  \bibinfo {author} {\bibfnamefont {J.}~\bibnamefont {Pogliano}},\ }\href@noop
  {} {\bibfield  {journal} {\bibinfo  {journal} {Proceedings of the National
  Academy of Sciences}\ }\textbf {\bibinfo {volume} {110}},\ \bibinfo {pages}
  {201311066} (\bibinfo {year} {2013})}\BibitemShut {NoStop}%
\bibitem [{\citenamefont {Lamsa}\ \emph {et~al.}(2016)\citenamefont {Lamsa},
  \citenamefont {Lopez-Garrido}, \citenamefont {Quach}, \citenamefont {Riley},
  \citenamefont {Pogliano},\ and\ \citenamefont {Pogliano}}]{lamsa2016}%
  \BibitemOpen
  \bibfield  {author} {\bibinfo {author} {\bibfnamefont {A.}~\bibnamefont
  {Lamsa}}, \bibinfo {author} {\bibfnamefont {J.}~\bibnamefont
  {Lopez-Garrido}}, \bibinfo {author} {\bibfnamefont {D.}~\bibnamefont
  {Quach}}, \bibinfo {author} {\bibfnamefont {E.~P.}\ \bibnamefont {Riley}},
  \bibinfo {author} {\bibfnamefont {J.}~\bibnamefont {Pogliano}}, \ and\
  \bibinfo {author} {\bibfnamefont {K.}~\bibnamefont {Pogliano}},\ }\href@noop
  {} {\bibfield  {journal} {\bibinfo  {journal} {ACS chemical biology}\
  }\textbf {\bibinfo {volume} {11}},\ \bibinfo {pages} {2222} (\bibinfo {year}
  {2016})}\BibitemShut {NoStop}%
\bibitem [{\citenamefont {Htoo}\ \emph {et~al.}(2019)\citenamefont {Htoo},
  \citenamefont {Brumage}, \citenamefont {Chaikeeratisak}, \citenamefont
  {Tsunemoto}, \citenamefont {Sugie}, \citenamefont {Tribuddharat},
  \citenamefont {Pogliano},\ and\ \citenamefont {Nonejuie}}]{htoo2019}%
  \BibitemOpen
  \bibfield  {author} {\bibinfo {author} {\bibfnamefont {H.~H.}\ \bibnamefont
  {Htoo}}, \bibinfo {author} {\bibfnamefont {L.}~\bibnamefont {Brumage}},
  \bibinfo {author} {\bibfnamefont {V.}~\bibnamefont {Chaikeeratisak}},
  \bibinfo {author} {\bibfnamefont {H.}~\bibnamefont {Tsunemoto}}, \bibinfo
  {author} {\bibfnamefont {J.}~\bibnamefont {Sugie}}, \bibinfo {author}
  {\bibfnamefont {C.}~\bibnamefont {Tribuddharat}}, \bibinfo {author}
  {\bibfnamefont {J.}~\bibnamefont {Pogliano}}, \ and\ \bibinfo {author}
  {\bibfnamefont {P.}~\bibnamefont {Nonejuie}},\ }\href@noop {} {\bibfield
  {journal} {\bibinfo  {journal} {Antimicrobial agents and chemotherapy}\
  }\textbf {\bibinfo {volume} {63}} (\bibinfo {year} {2019})}\BibitemShut
  {NoStop}%
\bibitem [{\citenamefont {Ojkic}\ \emph {et~al.}(2020)\citenamefont {Ojkic},
  \citenamefont {Lilja}, \citenamefont {Direito}, \citenamefont {Dawson},
  \citenamefont {Allen},\ and\ \citenamefont {Waclaw}}]{ojkic2020}%
  \BibitemOpen
  \bibfield  {author} {\bibinfo {author} {\bibfnamefont {N.}~\bibnamefont
  {Ojkic}}, \bibinfo {author} {\bibfnamefont {E.}~\bibnamefont {Lilja}},
  \bibinfo {author} {\bibfnamefont {S.}~\bibnamefont {Direito}}, \bibinfo
  {author} {\bibfnamefont {A.}~\bibnamefont {Dawson}}, \bibinfo {author}
  {\bibfnamefont {R.~J.}\ \bibnamefont {Allen}}, \ and\ \bibinfo {author}
  {\bibfnamefont {B.}~\bibnamefont {Waclaw}},\ }\href@noop {} {\bibfield
  {journal} {\bibinfo  {journal} {Antimicrobial Agents and Chemotherapy}\
  }\textbf {\bibinfo {volume} {64}} (\bibinfo {year} {2020})}\BibitemShut
  {NoStop}%
\bibitem [{\citenamefont {Wong}\ \emph {et~al.}(2021)\citenamefont {Wong},
  \citenamefont {Stokes}, \citenamefont {Cervantes}, \citenamefont {Penkov},
  \citenamefont {Friedrichs}, \citenamefont {Renner},\ and\ \citenamefont
  {Collins}}]{wong2021}%
  \BibitemOpen
  \bibfield  {author} {\bibinfo {author} {\bibfnamefont {F.}~\bibnamefont
  {Wong}}, \bibinfo {author} {\bibfnamefont {J.~M.}\ \bibnamefont {Stokes}},
  \bibinfo {author} {\bibfnamefont {B.}~\bibnamefont {Cervantes}}, \bibinfo
  {author} {\bibfnamefont {S.}~\bibnamefont {Penkov}}, \bibinfo {author}
  {\bibfnamefont {J.}~\bibnamefont {Friedrichs}}, \bibinfo {author}
  {\bibfnamefont {L.~D.}\ \bibnamefont {Renner}}, \ and\ \bibinfo {author}
  {\bibfnamefont {J.~J.}\ \bibnamefont {Collins}},\ }\href@noop {} {\bibfield
  {journal} {\bibinfo  {journal} {Nature Communications}\ }\textbf {\bibinfo
  {volume} {12}},\ \bibinfo {pages} {1} (\bibinfo {year} {2021})}\BibitemShut
  {NoStop}%
\bibitem [{\citenamefont {Mori}\ \emph {et~al.}(2017)\citenamefont {Mori},
  \citenamefont {Schink}, \citenamefont {Erickson}, \citenamefont {Gerland},\
  and\ \citenamefont {Hwa}}]{mori2017}%
  \BibitemOpen
  \bibfield  {author} {\bibinfo {author} {\bibfnamefont {M.}~\bibnamefont
  {Mori}}, \bibinfo {author} {\bibfnamefont {S.}~\bibnamefont {Schink}},
  \bibinfo {author} {\bibfnamefont {D.~W.}\ \bibnamefont {Erickson}}, \bibinfo
  {author} {\bibfnamefont {U.}~\bibnamefont {Gerland}}, \ and\ \bibinfo
  {author} {\bibfnamefont {T.}~\bibnamefont {Hwa}},\ }\href@noop {} {\bibfield
  {journal} {\bibinfo  {journal} {Nature communications}\ }\textbf {\bibinfo
  {volume} {8}},\ \bibinfo {pages} {1} (\bibinfo {year} {2017})}\BibitemShut
  {NoStop}%
\bibitem [{\citenamefont {Erickson}\ \emph {et~al.}(2017)\citenamefont
  {Erickson}, \citenamefont {Schink}, \citenamefont {Patsalo}, \citenamefont
  {Williamson}, \citenamefont {Gerland},\ and\ \citenamefont
  {Hwa}}]{erickson2017}%
  \BibitemOpen
  \bibfield  {author} {\bibinfo {author} {\bibfnamefont {D.~W.}\ \bibnamefont
  {Erickson}}, \bibinfo {author} {\bibfnamefont {S.~J.}\ \bibnamefont
  {Schink}}, \bibinfo {author} {\bibfnamefont {V.}~\bibnamefont {Patsalo}},
  \bibinfo {author} {\bibfnamefont {J.~R.}\ \bibnamefont {Williamson}},
  \bibinfo {author} {\bibfnamefont {U.}~\bibnamefont {Gerland}}, \ and\
  \bibinfo {author} {\bibfnamefont {T.}~\bibnamefont {Hwa}},\ }\href@noop {}
  {\bibfield  {journal} {\bibinfo  {journal} {Nature}\ }\textbf {\bibinfo
  {volume} {551}},\ \bibinfo {pages} {119} (\bibinfo {year}
  {2017})}\BibitemShut {NoStop}%
\bibitem [{\citenamefont {Basan}\ \emph {et~al.}(2020)\citenamefont {Basan},
  \citenamefont {Honda}, \citenamefont {Christodoulou}, \citenamefont
  {H{\"o}rl}, \citenamefont {Chang}, \citenamefont {Leoncini}, \citenamefont
  {Mukherjee}, \citenamefont {Okano}, \citenamefont {Taylor}, \citenamefont
  {Silverman} \emph {et~al.}}]{basan2020}%
  \BibitemOpen
  \bibfield  {author} {\bibinfo {author} {\bibfnamefont {M.}~\bibnamefont
  {Basan}}, \bibinfo {author} {\bibfnamefont {T.}~\bibnamefont {Honda}},
  \bibinfo {author} {\bibfnamefont {D.}~\bibnamefont {Christodoulou}}, \bibinfo
  {author} {\bibfnamefont {M.}~\bibnamefont {H{\"o}rl}}, \bibinfo {author}
  {\bibfnamefont {Y.-F.}\ \bibnamefont {Chang}}, \bibinfo {author}
  {\bibfnamefont {E.}~\bibnamefont {Leoncini}}, \bibinfo {author}
  {\bibfnamefont {A.}~\bibnamefont {Mukherjee}}, \bibinfo {author}
  {\bibfnamefont {H.}~\bibnamefont {Okano}}, \bibinfo {author} {\bibfnamefont
  {B.~R.}\ \bibnamefont {Taylor}}, \bibinfo {author} {\bibfnamefont {J.~M.}\
  \bibnamefont {Silverman}},  \emph {et~al.},\ }\href@noop {} {\bibfield
  {journal} {\bibinfo  {journal} {Nature}\ }\textbf {\bibinfo {volume} {584}},\
  \bibinfo {pages} {470} (\bibinfo {year} {2020})}\BibitemShut {NoStop}%
\bibitem [{\citenamefont {Bakshi}\ \emph {et~al.}(2021)\citenamefont {Bakshi},
  \citenamefont {Leoncini}, \citenamefont {Baker}, \citenamefont
  {Ca{\~n}as-Duarte}, \citenamefont {Okumus},\ and\ \citenamefont
  {Paulsson}}]{bakshi2021}%
  \BibitemOpen
  \bibfield  {author} {\bibinfo {author} {\bibfnamefont {S.}~\bibnamefont
  {Bakshi}}, \bibinfo {author} {\bibfnamefont {E.}~\bibnamefont {Leoncini}},
  \bibinfo {author} {\bibfnamefont {C.}~\bibnamefont {Baker}}, \bibinfo
  {author} {\bibfnamefont {S.~J.}\ \bibnamefont {Ca{\~n}as-Duarte}}, \bibinfo
  {author} {\bibfnamefont {B.}~\bibnamefont {Okumus}}, \ and\ \bibinfo {author}
  {\bibfnamefont {J.}~\bibnamefont {Paulsson}},\ }\href@noop {} {\bibfield
  {journal} {\bibinfo  {journal} {Nature Microbiology}\ }\textbf {\bibinfo
  {volume} {6}},\ \bibinfo {pages} {783} (\bibinfo {year} {2021})}\BibitemShut
  {NoStop}%
\end{thebibliography}%

\end{document}